\documentclass[a4paper,11pt]{article}
\pdfoutput=1 

\usepackage{jheppub} 
\usepackage{slashed} 
\usepackage[T1]{fontenc} 
\usepackage[utf8]{inputenc}

\def\eps{\epsilon}
\def\mmu2om2{\left (\frac{\mu^2}{m^2} \right )^{\!\eps}}

\def\spa#1.#2{\langle #1 #2\rangle}
\def\spb#1.#2{[ #1 #2]}
\def\spab#1.#2.#3{\langle #1 |#2| #3] }

\def\MS{{\overline{\rm{MS}}}\;}

\def\tcut#1{\tau_{#1}^{\rm{cut}}}

\title{Bottom-induced contributions to Higgs plus jet at next-to-next-to-leading order}

\author{Roberto Mondini}
\author{and Ciaran Williams}    

\affiliation{Department of Physics,\\University at Buffalo, The State University of New York, Buffalo
14260, USA}

\emailAdd{rmondini@buffalo.edu}
\emailAdd{ciaranwi@buffalo.edu}

\begin{abstract}{
We present a next-to-next-to-leading order (NNLO) QCD calculation of the bottom-induced contributions to the production of a Higgs boson plus a jet, i.e.~the process $p p \rightarrow H +j$ to $\mathcal{O}(y_b^2 \alpha_s^3)$. We work in the five-flavor scheme (5FS) in which the bottom quark mass is retained only 
in the coupling to the Higgs boson. Our calculation uses $N$-jettiness slicing to regulate infrared divergences, allowing for fully-differential predictions for collider observables. 
After extensively validating the methodology, we present results for the 13 TeV LHC. Our NNLO predictions show a marked improvement in the overall 
renormalization and factorization scale dependence, the latter of which proves to be particularly troublesome for 5FS calculations at lower orders.  
In addition, using the same methodology we present a NNLO computation of $b\overline{b} \rightarrow H$. Our results are implemented into MCFM.
 }
\end{abstract}

\begin{document} 
\maketitle
\flushbottom

\section{Introduction} 

Since its discovery nearly a decade ago~\cite{Aad:2012tfa,Chatrchyan:2012xdj}, the Higgs boson has become an established part of the particle physics landscape, and a significant amount of research effort has been devoted to the greater understanding of its properties. After the initial establishment of its intrinsic properties 
such as its CP and spin~\cite{Aad:2014xva,Khachatryan:2014jba},  the focus has shifted to obtaining precision measurements of the couplings of the Higgs boson to the other particles of the Standard Model (SM) as well as to itself. 
The Higgs boson self-coupling is a particularly pressing measurement to obtain, since it will allow for a more detailed study of the electroweak symmetry breaking potential in the Standard Model. Although the Higgs self-coupling 
is fully predicted from known parameters in the Standard Model, it is often sensitive to extensions of the SM (referred to as BSM) that change the nature of the electroweak potential. 
Similarly, it is also of vital importance to constrain the couplings of the Higgs boson to other particles in the Standard Model, namely the $W$ and $Z$ bosons and massive fermions. 
The LHC has made significant progress over the last decade~\cite{Aad:2019mbh,Sirunyan:2018koj}, and will continue to improve upon existing results over the forthcoming Run III. Further in the future, measurements 
with sub-percentage uncertainties will require a collider with upgraded capabilities, for which serious planning is now underway~\cite{Abada:2019lih,Contino:2016spe}. In all of these endeavors, precision predictions for differential distributions in the Standard Model are critical in order to avoid the situation in which theoretical uncertainties become the dominant  source of error.

Of the Higgs interactions, one of the most fascinating to study is the coupling between the Higgs boson and third-generation fermions, the top and bottom quarks and tau leptons. 
Among these, the bottom quark is unique, as there are two ways to gleam insights into its coupling: through Higgs boson production or its direct decay to the quarks themselves. 
 Given the large hierarchy in mass, the top quark dominates 
Higgs boson production at the LHC through the gluon-fusion mechanism and therefore probing the bottom Yukawa coupling through Higgs production is challenging. 
On the other hand, the SM Higgs boson copiously decays to bottom quarks with a branching fraction of around 50\%, and therefore the bottom Yukawa coupling dominates the Higgs decay width, propagating to all other (on-shell) measurements of the Higgs boson at the LHC. Determining the bottom-Higgs coupling as precisely as possible is thus an essential requirement of the future 
experimental high-energy physics program. For example, in extensions of the SM, extended Higgs sectors typically modify the coupling of the 125-GeV Higgs boson to up- and down-type fermions, and can 
lead to enhanced production cross sections~\cite{Cohen:2017rsk,Arcadi:2019lka,Dawson:2011pe,Dawson:2007ur}. 

Given the immense interest, there have been many theoretical studies of processes involving the Higgs boson and the bottom quark at hadron colliders. When making predictions at the LHC, one must first decide how to handle the mass of the bottom quark. Since the bottom quark is heavier than the proton, a natural choice is to keep the mass of the bottom quark in the calculation and exclude it from initial-state contributions. This scheme is known as the four-flavor scheme  (4FS), due to the number of active initial-state flavors in the proton.
The leading-order production mechanism at the LHC in the 4FS is thus the process $gg\rightarrow b\overline{b}H$ (plus a sub-dominant $q\overline{q}$-initiated contribution).
 The 4FS has the advantage 
that no approximations are made in regards to the kinematics, which is particularly helpful in relation to final-state bottom-quark tagging, since single $b$-tagged jets can be isolated 
without theoretical issues, i.e. no jet cuts are required even though the final state contains two partons. However, a major drawback of the 4FS scheme arises from the occurrence of large logarithms of the form  $\alpha_s \log (m_b^2/m_H^2)$. When computing cross sections as  perturbation 
series directly in $\alpha_s$, these logarithms induce large corrections and spoil the convergence of the series. One way to ameliorate this problem is to resum the logarithms when possible. Initial-state collinear logarithms can be
resummed into the parton distribution functions (PDFs). This introduces the five-flavor scheme (5FS) in which the bottom quark contributes to the initial-state PDFs, and the mass of the quark is neglected in the remaining 
kinematics. At leading order, Higgs boson production in the 5FS thus proceeds directly through bottom-quark fusion $b\overline{b}\rightarrow H$.  
It is worth noting that, since final-state collinear splittings $g\rightarrow b\overline{b}$ are not resummed, care must be taken in the 5FS when applying $b$-tagging requirements and comparisons are made to
experimental data (i.e.~one should try to remove jets arising from gluon splitting from the experimental analyses). 

Given this theoretical richness, there have been many detailed calculations, phenomenological studies, and comparisons of the 5FS and 4FS at various orders in $\alpha_s$, matching to parton-showers, and resummed predictions. We refer the interested reader to refs.~\cite{Dicus:1998hs,Balazs:1998sb,Campbell:2002zm,Maltoni:2003pn,Dawson:2003kb,Dawson:2004sh,Harlander:2010cz,Harlander:2011aa,Buehler:2012cu,Harlander:2012pb,Maltoni:2012pa,Wiesemann:2014ioa,Bonvini:2015pxa,Jager:2015hka,Forte:2016sja,Krauss:2016orf,Deutschmann:2018avk,H:2019nsw,Pagani:2020rsg,Grojean:2020ech} for further details. A particularly impressive calculation is the recent achievement of next-to-next-to-next-to-leading order (N3LO) accuracy for the total cross section in the 5FS~\cite{Duhr:2019kwi}, which has been subsequently matched to the NLO 4FS result~\cite{Duhr:2020kzd}.

A further intricacy relating the 5FS to the 4FS comes from the dependence on the factorization scale through the bottom-quark PDF in the 5FS. 
It was noted in the earliest calculations of bottom-quark fusion at NLO in the 5FS~\cite{Dicus:1998hs,Balazs:1998sb} that the higher-order corrections were large and resulted in cross sections 
that could have differences of an order of magnitude from the 4FS LO result. Detailed analysis in ref.~\cite{Maltoni:2003pn} (following the arguments of ref.~\cite{Plehn:2002vy}) illustrated that the 
choice of a central factorization scale of $m_H$ was too high for the process, and that a scale of around $m_H/4$ was more appropriate to ensure the reliability of collinear factorization. 
Predictions made with a central factorization scale choice in this region showed much better perturbative convergence, and a broader compatibility with the 4FS result.
Nevertheless, a strong dependence on the unphysical factorization scale should be seen as a negative feature of the 5FS when used at LO and NLO. Subsequent NNLO and N3LO predictions for bottom-quark fusion~\cite{Buehler:2012cu,Harlander:2012pb,Duhr:2019kwi} show a significant reduction of this problem and motivate our computation of $H+j$ in the 5FS at NNLO.

On the Higgs boson decay side the theoretical situation is also under good control. Again, there are two scheme possibilities which can be considered as the decay versions of those discussed above. 
A commonly-used approximation is to retain the mass of the bottom quark only in the coupling to the Higgs boson (the decay equivalent of the 5FS), and there have been many detailed theoretical studies of this process (see e.g.~\cite{Chetyrkin:1996sr,Anastasiou:2011qx,DelDuca:2015zqa}), which is now known up to $\mathcal{O}(\alpha_s^4)$ inclusively~\cite{Baikov:2005rw} and at N3LO differentially~\cite{Mondini:2019gid}. Increasing the complexity of the calculation, one can include the mass of the bottom quarks fully (the 4FS equivalent), and in this setup recent calculations 
have pushed the accuracy to NNLO for fully-differential predictions~\cite{Bernreuther:2018ynm,Behring:2019oci}. 

The continuing maturation of the experimental analyses at the LHC has had a twofold impact on Higgs boson studies. Firstly, the increased statistics and 
precision have allowed for an extensive range of Higgs boson observables to be studied, including Higgs-plus-multiple-jet production (see e.g.~\cite{Aaboud:2018xdt,Sirunyan:2018sgc}). Secondly, comparison of 
data to theory has highlighted the need for increased precision on the theoretical front, emphasizing the importance of NNLO predictions 
in QCD. Over the last few years significant progress has been made, resulting in several independent calculations of Higgs-plus-jet at NNLO in the effective 
field theory (EFT) in which the top quark is integrated out~\cite{Chen:2014gva,Boughezal:2015dra,Boughezal:2015aha,Chen:2016zka,Chen:2018pzu,Chen:2019wxf,Campbell:2019gmd}. Impressively, in ref.~\cite{Cieri:2018oms}
the accompanying jet was integrated out of the calculation, allowing for a computation of the $pp\rightarrow H+X$ differential cross section at N3LO accuracy.

Our aim in this paper is to provide a similar level of theoretical accuracy for the bottom quark-initiated contribution to $H+j$ as it is present for the dominant EFT production mechanism. 
In order to do so, we will work in the 5FS and treat the bottom quark as massless everywhere except in the coupling between the Higgs boson and the bottom quarks. 
The different calculations of $H+j$ at NNLO available in the literature used a variety of infrared-regulating techniques. For brevity we do not describe them all in detail here, but focus on the 
pieces pertinent to this paper. One of the initial calculations of $H+j$~\cite{Boughezal:2015aha} used a non-local slicing procedure (based on the event shape $N$-jettiness). The results obtained in this paper seemed to be in conflict with those obtained in other studies based upon local subtraction techniques. A detailed study in ref.~\cite{Campbell:2019gmd} showed that the two methodologies do indeed produce the same results, but that power corrections arising from the approximate form of the factorization formula used in the slicing procedure must be handled carefully. In this paper we will use the same methodology and follow the same techniques as shown in ref.~\cite{Campbell:2019gmd} to control and estimate the remaining power corrections. 
In ref.~\cite{Campbell:2019gmd} the $gg$ channel was shown to have the worst power corrections for EFT $H+j$ production. Thankfully for our calculation,  this channel does not appear at leading order and we therefore
expect power corrections to be easier to control. 

Our paper proceeds as follows. In section~\ref{section:calc} we present a brief overview of the technical details of our calculation, while section~\ref{section:val} discusses its validation. We present results for the 13 TeV LHC in section~\ref{section:results}, and finally we draw our conclusions in section~\ref{section:conc}.

\section{Calculation} 
\label{section:calc}

\subsection{General overview} 

The primary focus of this paper is the calculation of the NNLO QCD corrections to the bottom-induced contributions to Higgs plus one jet at the LHC.
Given its phenomenological relevance and role as a check of our calculation, we will also present results for the bottom quark fusion process at NNLO (i.e.~bottom-induced contributions to Higgs plus zero jets).

\begin{figure}
\begin{center}
\includegraphics[width=11cm]{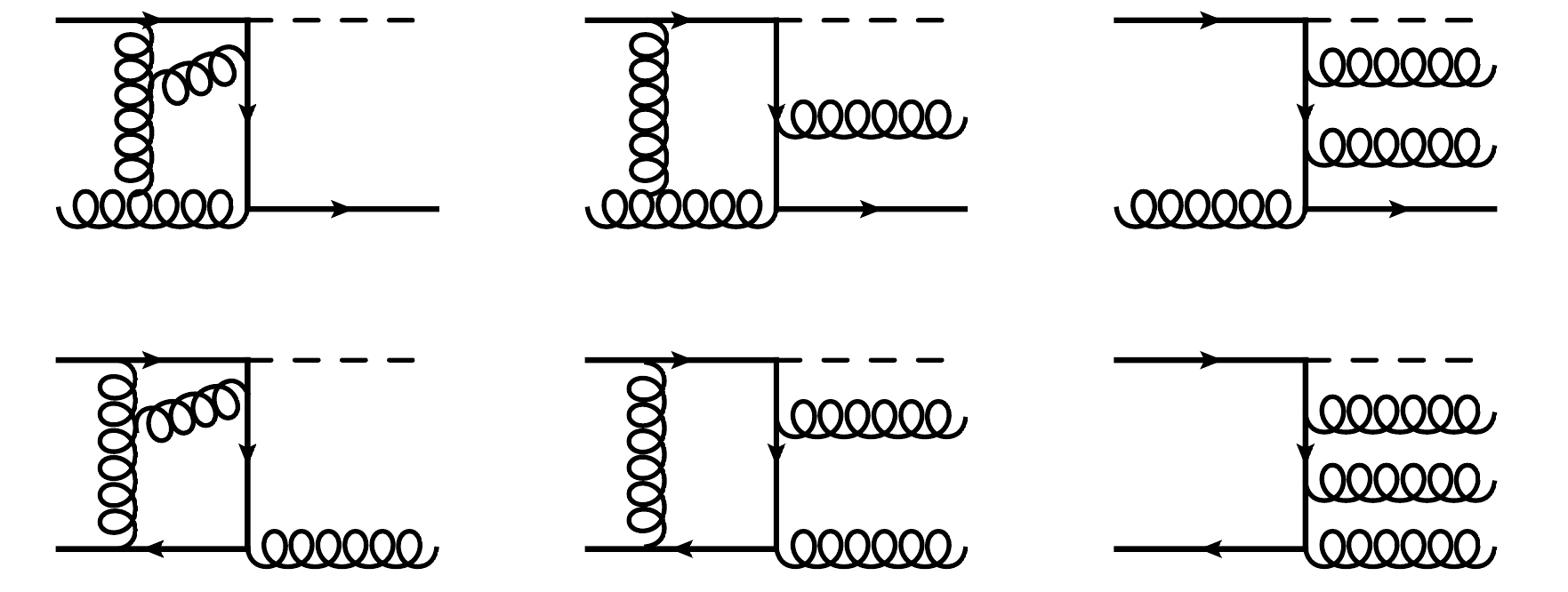}
\caption{Top row: representative Feynman diagrams in the 5FS which contribute to the process $pp\rightarrow H+j$ at NNLO, regardless of $b$-tagging requirements. Bottom row: representative Feynman diagrams in the 5FS which contribute to the process $pp\rightarrow H+j$ at NNLO, but fail $b$-tagging requirements.}
\label{fig:5fs}
\end{center}
\end{figure}
As discussed in the introduction, the most important theoretical choice when considering bottom-quark processes at hadron colliders is how to treat $m_b$, i.e.~whether to work in the 5-flavor (5FS) 
or 4-flavor (4FS) scheme.  In this paper we work in the five-flavor scheme, which will allow us to extend the 
computation to NNLO accuracy. Representative Feynman diagrams relevant for our calculation at this order are shown in fig.~\ref{fig:5fs}.
We recall that in the 5FS the bottom quark mass is taken to zero and bottom quarks have a non-zero 
contribution to the PDFs. At first glance, the 5FS scheme may appear not to be useful for computing $H+b$ related processes, since by setting the bottom quark 
mass to zero the bottom Yukawa coupling should also be taken to zero. In order to circumvent this problem, we work in the mixed-renormalization scheme, in which the bottom quark Yukawa is taken in the 
$\MS$ scheme, and the bottom quark mass, used in propagators and in the relativistic kinematics, is taken in the on-shell scheme. This scheme allows one to take the limit $m_b^{\rm{OS}} \rightarrow 0$ 
while keeping the Yukawa coupling non-zero. This scheme has two advantages in QCD calculations. Firstly, it allows for a robust definition of the 5FS for $H+b$ amplitudes. Secondly, by evolving the scale in 
the running Yukawa coupling to $\mu_R$ (i.e.~$m_H$), one avoids large logarithms which arise in the OS scheme at higher orders, and as a result the perturbative corrections are under better control. Downsides of the mixed scheme 
include breaking the relationship between the OS mass and the $\MS$ mass~\cite{Chetyrkin:1999qi,Melnikov:2000qh} and an inability to consistently renormalize higher-order corrections in the electroweak coupling~\cite{Pagani:2020rsg}. 
Nevertheless, the reduction in sensitivity to collinear initial-state logarithms (at the cost of a strong dependence on the factorization scale at LO), and the ability to pursue higher-order corrections, renders the 5FS along with the mixed renormalization scheme a very useful theoretical construct for LHC computations. 

\subsection{Technical details}

For the  bottom-induced $H+j$ process at NNLO, three phase-space topologies contribute (see fig.~\ref{fig:5fs}), corresponding to the double-virtual, real-virtual, and 
double-real corrections to the underlying LO topology. UV and IR singularities are present at this order and must be appropriately renormalized and regulated. We describe the calculation of the various 
UV-renormalized matrix elements for each phase-space configuration in ref.~\cite{Mondini:2019vub} for the decay $H\rightarrow b\overline{b} j$ at NNLO. This leaves the discussion 
of the IR regulation, which is different from that described in ref.~\cite{Mondini:2019vub} due to the LHC kinematics. 

In order to regulate the IR divergences present at this order we use the $N$-jettiness slicing approach~\cite{Gaunt:2015pea,Boughezal:2015dva}. 
This method has become an established technique for evaluating NNLO cross sections involving final-state jets at the 
LHC~\cite{Boughezal:2015dva,Boughezal:2015ded,Campbell:2016lzl,Campbell:2019gmd}, and we provide a brief overview in this section. 
The central idea is to separate the (differential) cross section of a process into two pieces, 
\begin{eqnarray}
\sigma^{\rm{NNLO}}= \sigma(\tau_N \le \tcut {n_j}) + \sigma(\tau_N  > \tcut {n_j}) \, ,
\end{eqnarray}
where the variable $\tau_N $ is the $N$-jettiness variable~\cite{Stewart:2010tn}. For our 1-jet example, this variable is defined as 
\begin{eqnarray}
{\tau_{1}}  = \sum_{m} {\rm{min}}_i \,\frac{2 p_m\cdot k_i}{P_i} \, ,
\end{eqnarray}
where $\{p_m\}$ is the set of all partonic momenta in an event, while $\{k_i\}$ are the momenta of the two incoming beams and the hardest jet present in the event (after clustering). The quantity $P_i$ is a somewhat arbitrary choice of hard scale, and in our calculation we take $P_i = 2E_i$ (known as the geometric measure ~\cite{Jouttenus:2011wh,Jouttenus:2013hs}).
The above-cut term $\sigma(\tau_N  > \tcut {n_j})$ has sufficiently large value of the $N$-jettiness variable to have at most one unresolved parton, and therefore corresponds to a NLO 
computation of the cross section with an additional parton present. 
The below-cut  term $\sigma(\tau_N \le \tcut {n_j})$ contains all of the double-unresolved 
limits at NNLO. 
However, in the limit $\tcut1 \rightarrow 0$ the cross section can be approximated using the following factorization theorem from Soft-Collinear Effective Field Theory (SCET):
\begin{eqnarray}
\sigma(\tau \le \tcut {n_j}) = \int_0^{\tcut {n_j}} d\tau \; \left(\mathcal{S} \otimes  \prod_{i=1}^{n_{\rm{j}}} \mathcal{J}_i  \otimes  \prod_{a=1,2} \mathcal{B}_a   \otimes \mathcal{H}\right) + \mathcal{F}( \tcut {n_j}),
\label{eq:scettau}
\end{eqnarray}
where in our case $n_{{j}}=1$. The above equation is valid up to power corrections (denoted by the $ \mathcal{F}(\tcut {n_j})$ term), which vanish in the limit $\tcut {n_j} \rightarrow 0$. 
At NLO the leading power corrections are well described by the form $\tcut {n_j} \log  (\tcut {n_j}/Q)$, and at NNLO the leading power corrections have the form  $\tcut {n_j} \log^3 (\tcut {n_j}/Q)$ (where in both cases $Q$ is the hard 
scale associated with the process).
The general terms that enter the SCET factorization theorem are the soft ($\mathcal{S}$), jet  ($\mathcal{J}$), and beam ($\mathcal{B}$) functions, for which calculations accurate to $\mathcal{O}(\alpha_s^2)$ needed for our calculation can be found in refs.~\cite{Campbell:2017hsw,Boughezal:2015eha,Becher:2010pd,Becher:2006qw,Gaunt:2014xga,Gaunt:2014cfa}.

There are several alternative choices~\cite{Campbell:2016lzl,Campbell:2019gmd} one can make when applying the jettiness-slicing method. Firstly, one can choose whether to work with a fixed version of $\tcut {1}$, in which all events are 
compared to a given energy scale, or with a dynamical definition, in which the final-state kinematics (of the clustered system) generates different $\tcut {1}$ values for each phase-space point. 
Typically, for 1-jet NNLO processes it is more prudent to use the latter option. Since power corrections are sensitive to the overall hardness of the system through the expansion parameter $\tcut{1}/Q$, very energetic jets have suppressed power corrections. By using a fixed $\tcut {1}$, the calculation for these terms includes points that are very soft and collinear (relative to the hard scale), resulting in large Monte Carlo uncertainties and code instabilities. On the other hand, using a dynamic $\tcut {1}$ ensures a more relaxed $\tcut {1}$ for more energetic jets, reducing this problem and producing more stable results, without increasing the impact of unwanted power corrections.

In order to obtain the remaining process-specific hard function ($\mathcal{H})$ appearing in  eq.~\eqref{eq:scettau}, we use our double-virtual calculation for the decay amplitude $H\rightarrow b\overline{b}g$ presented in 
ref.~\cite{Mondini:2019vub}\footnote{See also ref.~\cite{Ahmed:2014pka}.}. The result for LHC kinematics is obtained by performing the relevant crossing, moving the desired final-state partons to the initial state. 
In practical terms, this involves taking the appropriate analytic continuation of the various harmonic polylogarithms that appear in the virtual amplitudes as described in section 4 of ref.~\cite{Gehrmann:2002zr}.
After crossing the relevant final-state partons to the initial state, we have checked that our results have the correct factorization 
properties in the relevant soft and collinear limits~\cite{Li:2013lsa,Badger:2004uk}, finding excellent agreement. 

\subsection{Matching to the EFT} 

The Standard Model does not allow for the consideration of the impact of a single fermion generation in isolation. For the purposes of this calculation, in order to completely specify our theoretical framework we must also address the role of the top quark in the computation. This is because at $O(\alpha_s^3)$ the cross section becomes sensitive to the presence of the top induced production. 
One must therefore specify whether one works in the effective field theory or full Standard Model. Precision calculations in the full Standard Model are made considerably more difficult by the presence of the additional mass scale and are currently known to NLO accuracy for $H+j$~\cite{Jones:2018hbb}. On the other hand, NNLO predictions are available in the EFT~\cite{Chen:2014gva,Boughezal:2015dra,Boughezal:2015aha,Chen:2016zka,Chen:2018pzu,Chen:2019wxf,Campbell:2019gmd}.
Typically, in EFT calculations the top mass effects are included via a rescaling of the cross section by those computed in the full theory at lower orders.

For $H+j$ in the 5FS at $\mathcal{O}(\alpha_s^3)$ accuracy there are two contributions, the pure bottom-induced and the LO top-induced piece, and since the 
top-induced contribution is leading order one could easily consistently work in the full SM or the EFT. However, it is known that the higher corrections to the EFT pieces 
are large, and therefore including only the LO piece makes little phenomenological sense. Our strategy in this paper is to ignore the top-induced pieces altogether and 
focus only on the technical aspects of the NNLO calculation of the bottom-induced contribution. In order to obtain reliable phenomenological predictions at ``NNLO'', one would therefore wish to 
combine the $\mathcal{O}(\alpha_s^3 y_b^2)$ pieces with the $\mathcal{O}(\alpha_s^5)$ EFT results. To avoid having the bottom-induced component be entirely overwhelmed by 
the EFT piece, one would also wish to apply $b$-tagging requirements (and consider other sources of Higgs plus heavy flavor~\cite{Pagani:2020rsg} arising from VBF and $VH$ processes).
We postpone such detailed phenomenology study to a future publication.  Additionally, we note that when working in the EFT the bottom Yukawa coefficient is matched to that of the full SM as follows, 
\begin{eqnarray}
y_b^{\rm{EFT}} = y_b^{SM}\left(1+ \left(\frac{\alpha_s}{\pi}\right)^2 \Delta^{(2)}_F+\mathcal{O}(\alpha_s^3)\right) \, ,
\end{eqnarray}
where 
\begin{eqnarray}
\Delta^{(2)}_F =\left(\frac{5}{18} - \frac{1}{3}\log\frac{\mu^2}{m_t^2}\right) \, .
\end{eqnarray}
This means that, when working in the context of the EFT, we should include a term proportional to $\Delta_F^{(2)}$ multiplying our LO  predictions. 
In this paper we remain agnostic to the exact implementation of the top quark and therefore choose to present results in terms of the unmatched $y_b^{SM}$. 
The impact of adjusting the coefficient to $y_b^{\rm{EFT}}$ is a rather small (sub-percentage) effect and does not affect the conclusions presented in this paper.

Finally, we note that there are interference terms between the top quark (or EFT) initiated contributions to $H+j$ and the bottom-induced contributions. 
This interference requires a helicity flip in order to be non-zero, inducing an overall scaling of the form $y_b^{\MS} m_b^{\rm{OS}}$ (since the helicity flip is a kinematic mass). As a result, the interference vanishes in the 5FS.
However, there is an ambiguity in the mixed-renormalization scheme which renders this argument not quite complete, since one can relate the OS mass to the $\MS$ mass changing the scaling to $y_b^{\MS} m_b^{\MS} \propto (y_b^{\MS})^2$  and {\it{then}} take the limit $m_b^{\rm{OS}} \rightarrow 0$ to approach the 5FS.  At ``LO'' in the interference, $\mathcal{O}(\alpha_s^2 y_b y_t)$, such a procedure is well-defined since the interchange of the mass schemes is trivial. However, at higher orders this procedure is much more delicate due to the presence of 
IR logarithms in $m_b^{\rm{OS}}$, and rich UV structure. Very recently, this limit was studied in the context of extracting a sensible result at NLO in the interference for Higgs-plus-charm production~\cite{Bizon:2021nvf} (where the even 
larger hierarchy between $y_c$ and $y_t$ makes these terms more important). 
In addition, these pieces were studied in ref.~\cite{Mondini:2020uyy} for the decay of $H\rightarrow b\overline{b}$ and $H\rightarrow c\overline{c}$ at $\mathcal{O}(\alpha_s^3)$ in the ``4FS'' in which the mass was fully retained.
In keeping our focus on the technical aspects of the NNLO computation, and being agnostic regarding the top quark implementation, in this paper we take the first limit, in which the interference is set to zero. However, when pursuing a full LHC phenomenology study and given the size of $y_t$, we advocate including the term using the limit extraction in which the helicity flip mass is coverted into the $\MS$ mass prior to taking the limit. We leave this to a more detailed future study. 

\section{Validation}
\label{section:val}

The calculation described in the previous section has been implemented into the Monte Carlo code MCFM~\cite{Campbell:1999ah,Campbell:2011bn,Campbell:2015qma,Boughezal:2016wmq}.
We make extensive use of the code's ability to handle processes involving a final-state jet at NNLO, and particularly important for this paper are the MCFM developments outlined in refs.~\cite{Campbell:2019gmd,Campbell:2016lzl}.
This section details the various checks we have performed on our computation (in addition to the analytic soft and collinear checks previously mentioned). 

\subsection{$ b\overline{b} \rightarrow H$ at NNLO} 

We validate our calculation of $b\overline{b} \rightarrow H$ at NNLO by comparing our results to those known in the literature. This process has been well studied and public codes are available for the the computation of cross sections at NNLO accuracy. We use the {\tt{SusHi}} framework~\cite{Harlander:2012pb,Harlander:2003ai}, which can compute a variety of Higgs production cross sections at NNLO accuracy in the SM and its supersymmetric extensions. In this comparison we run MCFM with parameters set to match the default implementation in {\tt{SusHi}}. 
We use the MMHT14~\cite{Harland-Lang:2014zoa} PDF sets, (taking the NNLO set for all predictions) and the following setup for our comparison:
\begin{center}
$\sqrt{s} = 13$ TeV,      $\mu_R/2=\mu_F=m_H$. 
\end{center}
As discussed in previous sections, the choice of a factorization scale around the Higgs boson mass is somewhat of a problem for phenomenology, since the perturbation theory is subject to large corrections. However, in this case the large corrections act in our favor when attempting to validate our implementation, since the larger NNLO coefficient allows us to separate scales associated with the pure coefficient, power-suppressed corrections, and numerical Monte Carlo uncertainties. With the parameter choices listed above, the values of the bottom quark mass used in the Yukawa couplings are $m^{\MS}_b(\mu_R) = \{ 2.868, 2.666, 2.645\} $ GeV at LO, NLO, and NNLO respectively. By matching the MCFM parameters to these values we observe excellent agreement at NLO: the result from {\tt{Sushi}} is 701.97 fb, while MCFM gives 701.95 fb. 
\begin{figure}
\begin{center}
\includegraphics[width=14cm]{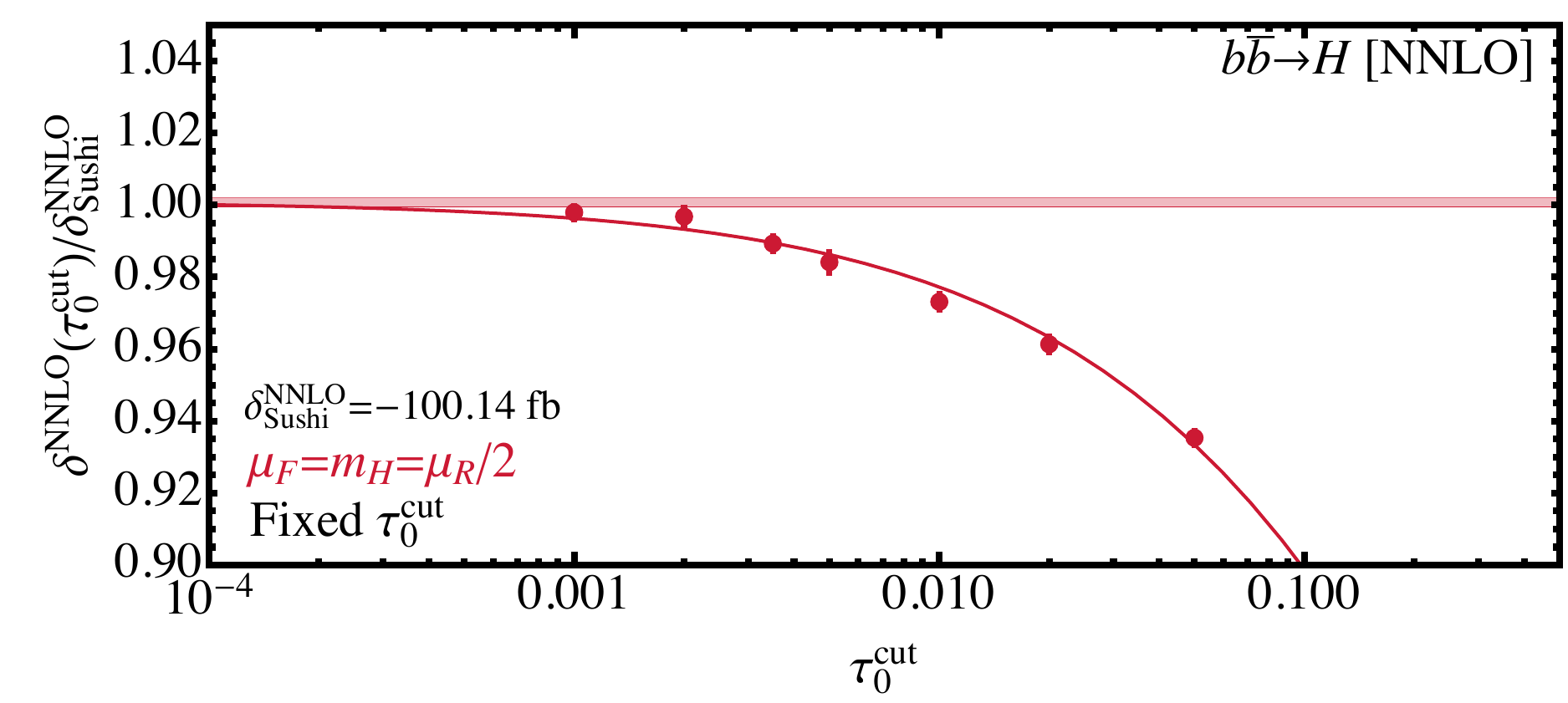}
\caption{Comparison of the NNLO coefficient obtained with MCFM and the equivalent prediction from {\tt{SusHi}}. Shown are the results obtained for a selection of
choices of the slicing parameter $\tau_0^{\rm{cut}}$ and a two-parameter fit to the power corrections. The shaded band corresponds to the uncertainty on the extracted $\tau_0^{\rm{cut}}\rightarrow 0$ 
limit from the fit.}
\label{fig:taubbHSus}
\end{center}
\end{figure}
We then proceed to compare our result for the NNLO coefficient in fig.~\ref{fig:taubbHSus}. Here we use the $\tau_0$ parameter in the laboratory (unboosted) frame and present results over the range 
$0.001 \le \tau_0^{\rm{cut}} \le 0.05$ GeV. As is by now well known in the literature, the leading power corrections at NNLO can be described parametrically as follows, 
\begin{eqnarray}
\delta^{\rm{NNLO}}(\tau_0^{\rm{cut}}) = \delta^{\rm{NNLO}}_0 + c_0 \left(\frac{\tau_0^{\rm{cut}}}{Q}\right)\log^3{\left(\frac{\tau_0^{\rm{cut}}}{Q}\right)} +\dots \, ,
\end{eqnarray}
where the ellipses indicate sub-leading contributions of the form $\tau \log^2\tau$ etc., and $Q$ is a hard scale associated with the process (e.g.~$m_H$). 
The residual power corrections in our results are clearly well described by this parametric form, and by fitting our results accordingly we are able to simultaneously extract the 
coefficient in the limit $\tau_0^{\rm{cut}} \rightarrow 0$ and parametrize  the residual impact of power corrections present in calculations with non-zero $\tau_0^{\rm{cut}}$. 
By fitting our results in this way we determine 
\begin{eqnarray}
\delta^{\rm{NNLO}}_0 = -100.20 \pm 0.13 \; {\rm{fb}} \, ,
\end{eqnarray}
which is in excellent agreement with the coefficient obtained from {\tt{Sushi}}, $\delta^{\rm{NNLO}}_{\rm{Sushi}} = -100.14$ fb. 
Our results clearly show that for $\tau_{0}^{\rm{cut}}$ in the region $10^{-3}$ GeV the residual power corrections are significantly less than 1\% of the NNLO coefficient, and subsequently per-mille level relative to the total physical prediction. In addition to the detailed comparison described above, we have performed a similar fit to our calculation with the canonical scale choice of $\mu_R = m_H = 4\,\mu_F$, obtaining $\delta_0^{\rm{NNLO}} = 18.39 \pm 0.18$ fb, which is again in excellent agreement with the result obtained from {\tt{SusHi}}, 18.52 fb. 

\subsection{$b\overline{b} \rightarrow H+j$ at NLO} 

Before studying the slicing dependence of the main result of our paper, the bottom Yukawa contributions to $H+j$ at NNLO, we study the process at NLO. 
The primary area of interest is to study the different options for the definition of $\tcut1$ and their associated asymptotic regions of validity. 
In order to test the various ingredients of our calculation, we begin by computing cross sections for $H+j$ at NLO using the different IR-regulating prescriptions described in the previous sections. 
For these comparisons we use the following setup: 
\begin{center}
$\sqrt{s} = 13$ TeV,      $\mu_R=4\mu_F=m_H$ \\
  $p_T^{j} > 30$ GeV, $|\eta_{j}| < 4.5$,
\end{center}
with jets clustered using the anti-$k_T$ algorithm with $R=0.4$. Additionally, we will briefly study the power corrections with the more central jet requirement of $|\eta_{j}| < 2.5$. No cuts are implemented on the Higgs boson.  We use the MMHT14~\cite{Harland-Lang:2014zoa} PDF sets and for simplicity we use the NNLO PDF sets for all predictions in this section. Consequently, $\alpha_s$ and $m^{\MS}_b(\mu_R)$ are evaluated 
using the three-loop running (implemented into MCFM using the results of {\tt{RunDec}}~\cite{Chetyrkin:2000yt}). We take as an input $m_b^{\MS}(m_b) = 4.18$ GeV, such that  $m_b^{\MS}(m_H) = 2.793$ GeV. 
With our central scale choice and the parameters described above, the LO cross section is 92.61 fb.

\begin{figure}
\begin{center}
\includegraphics[width=7.5cm]{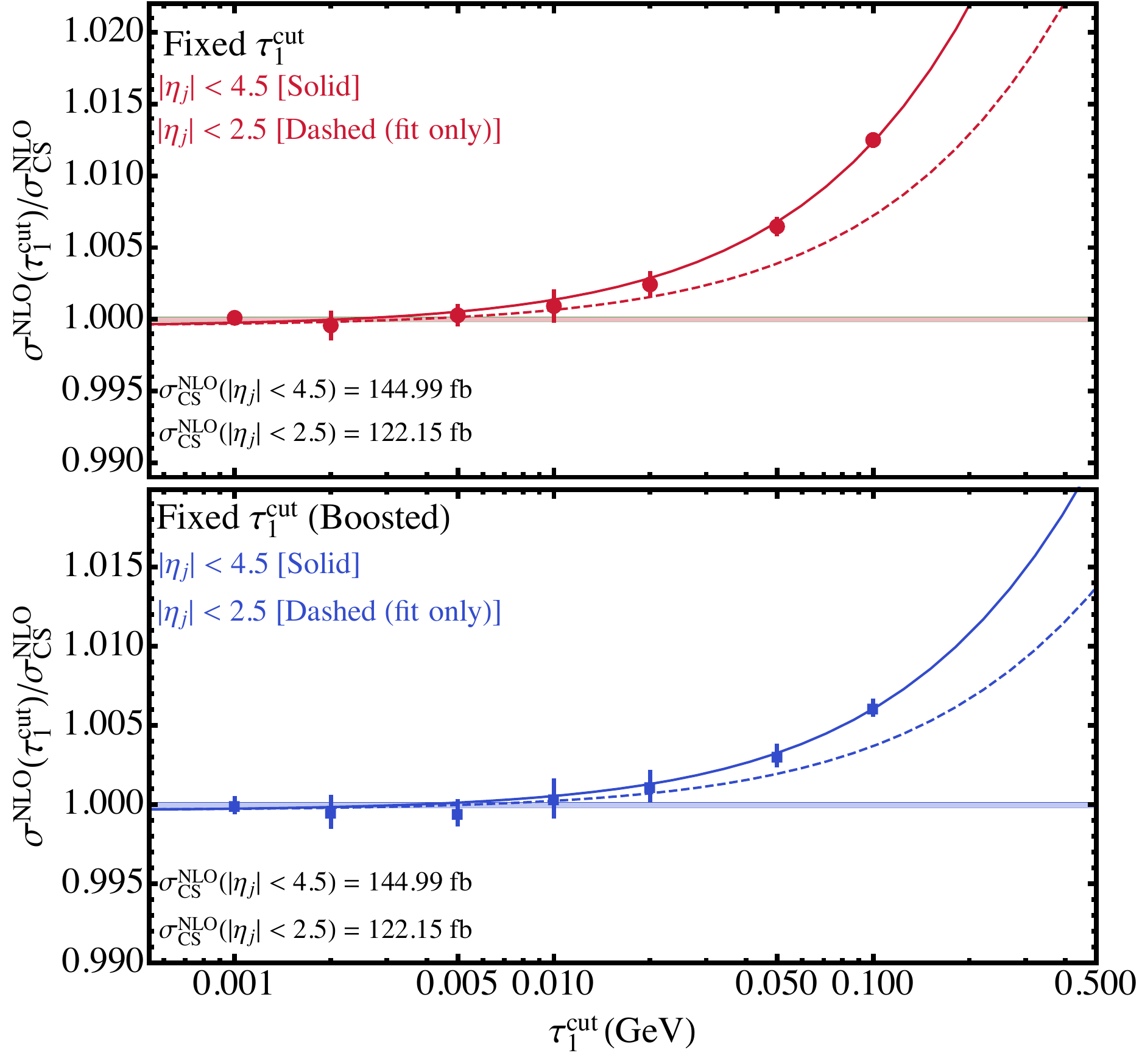}
\includegraphics[width=7.5cm]{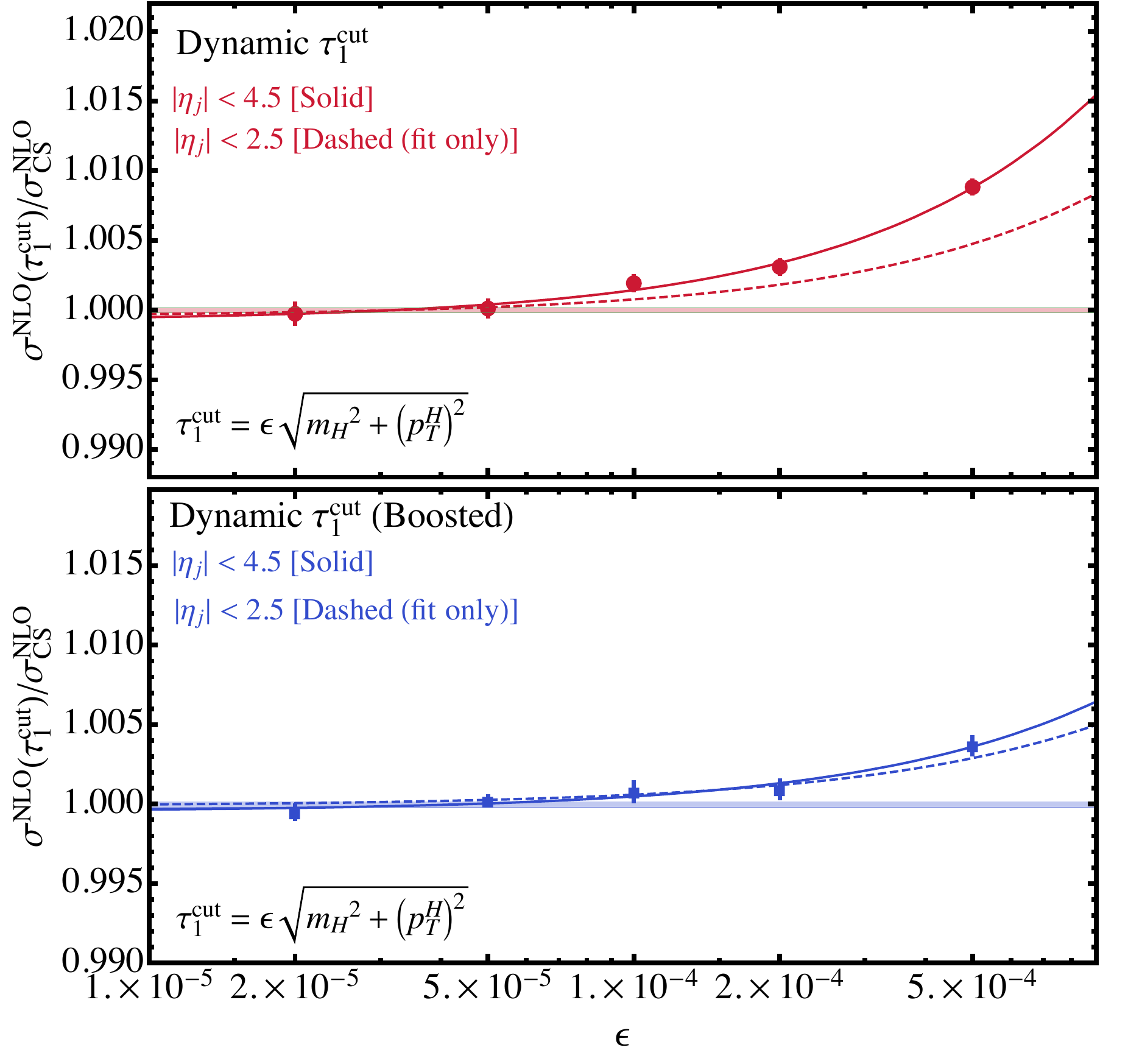}
\caption{A comparison of different definitions of the jettiness-slicing parameter for the NLO predictions of $H+j$.  The figures on the left use a
 fixed definition of $\tcut1$, while those on the right use a dynamic version. The upper panels show the result in the laboratory frame whereas the 
 lower panels evaluate the cut in the rest frame of the $H+j$ system. }
\label{fig:nloh1jcomp}
\end{center}
\end{figure}

Next, we turn our attention to validating the NLO cross section, which we have computed using dipole subtraction~\cite{Catani:1996vz} and the jettiness-slicing approach. As part of the validation of the 
dipole calculation we have checked the (in)dependence of our result on the unphysical $\alpha$ parameters \cite{Nagy:1998bb}, which control the amount of non-singular phase space utilized in the dipole subtractions. 
Using the dipole method and the parameter choices above, the corresponding NLO cross section is 144.98 fb. We now consider the various implementations of the jettiness-slicing method. 
Our results are presented in fig.~\ref{fig:nloh1jcomp}, where the panels on the left side show the ratio of the cross section obtained using a fixed value for $\tcut1$ to the dipole result, for both the boosted and traditional definitions. 
The data points on the figure show the results obtained with the full phase-space cuts described above as well as a fit to the data of the form 
\begin{eqnarray}
 \delta^{\rm{NLO}}_{\tau} = \delta^{\rm{NLO}}_0 + c_0 \left(\frac{\tcut1}{Q}\right) \log{\left(\frac{\tcut1}{Q}\right)}.
\end{eqnarray}
In order to quantify the impact of forward radiation on the power corrections we additionally show a fit to similar results obtained with a tighter jet requirement $|\eta_{j}| < 2.5$, although for readability we suppress the Monte Carlo output. 
The difference between the solid and dashed lines on the figure is therefore indicative of the sensitivity of the power-suppressed terms to the presence of forward jets.
The panels on the right side of fig.~\ref{fig:nloh1jcomp} show the same cross section ratios, computed using a dynamic version of $\tcut1$, which we define as 
\begin{eqnarray}
\tau_1^{\rm{cut}} = \epsilon \sqrt{m_H^2 + (p_T^H)^2} \, .
\end{eqnarray}
As before, results are evaluated in both the laboratory and boosted frames. The corresponding fit  for this setup is as follows,
 \begin{eqnarray}
 \delta^{\rm{NLO}}_{\epsilon} = \delta^{\rm{NLO}}_0 + c_0 \,\epsilon \log{\epsilon}. 
 \end{eqnarray}
We observe the same pattern for the impact of the power corrections as reported in ref.~\cite{Campbell:2019gmd}.  By evaluating in the boosted frame, the size of the power corrections is significantly reduced~\cite{Moult:2016fqy}, especially when the phase space includes contributions from regions in which the jet has larger pseudo-rapidity $(|\eta_{j}| > 2.5)$. Using the dynamic version of $\tcut1$ also results in smaller power corrections, and in particular the boosted-dynamic definition 
is the least sensitive to power corrections. We therefore employ the boosted-dynamic version of the slicing in our subsequent studies at NNLO.

\subsection{$b\overline{b} \rightarrow H+j$ at NNLO} 

\begin{figure}
\begin{center}
\includegraphics[width=15cm]{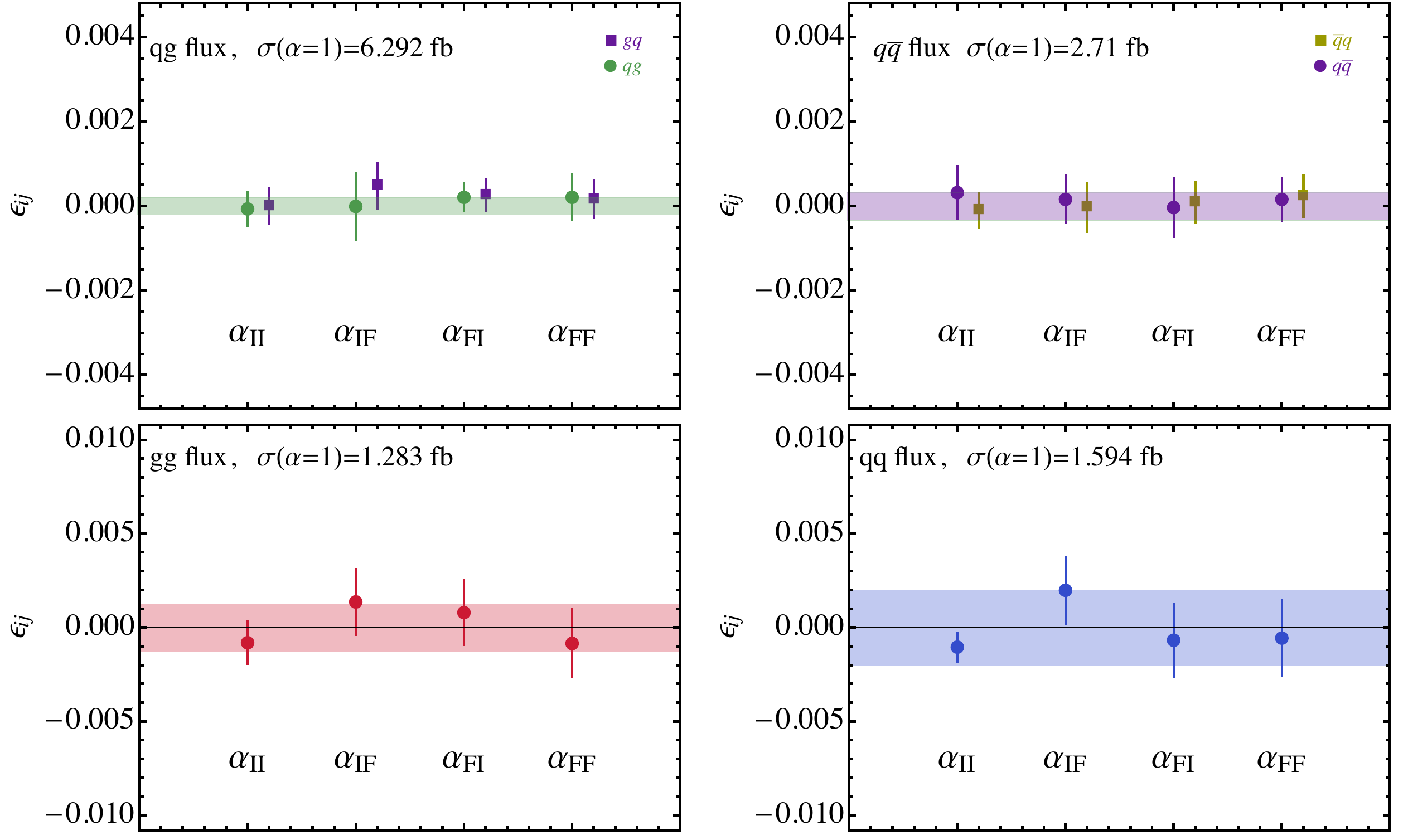}
\caption{The independence on the $\alpha$ parameter for a selection of partonic configurations for bottom-induced $H+2j$ production. The shaded band indicates the uncertainty on the $\alpha =1$ prediction.
The most intricate 
$qg$ and $q\overline{q}$ channels, which receive contributions from both $b\overline{b}ggH$ and four-quark amplitudes, have been computed with greater Monte Carlo statistics.}
\label{fig:H2jalpha}
\end{center}
\end{figure}

In this section we discuss the validation of our primary result, the NNLO predictions for the bottom-induced contributions to $H+j$. 
We begin by presenting a check of the $H+2j$ NLO result, which forms the above-cut piece of our NNLO prediction. 
As before, we check the $\alpha$ (in)dependence of the calculation, for which results are presented in fig.~\ref{fig:H2jalpha}.
These predictions were obtained using the NNLO CT14 PDF set~\cite{Dulat:2015mca}, $\mu_R = \mu_F = m_H$, and the two-loop running of the bottom Yukawa coefficient. 
By comparing results at sub per-mille level accuracy we are able to rigorously test the cancellation of IR singularities at one loop 
and the subsequent cancellation of dipole-related terms from the real-virtual and double-real contributions at NNLO. 
Following the notation of ref.~\cite{Campbell:2019gmd}, we define 
\begin{eqnarray}
\epsilon_{ab} = \frac{\sigma(\alpha_{ab} = 1) - \sigma(\alpha_{ab} = 10^{-2} )}{\sigma(\alpha_{ab} = 1) } \, ,
\end{eqnarray}
where the indices $a$ and $b$ correspond to either initial- $(I)$ or final- $(F)$ state dipoles. Fig.~\ref{fig:H2jalpha} illustrates that our results are insensitive 
to the choice of the $\alpha$ parameter at the level of $\epsilon \sim 0.0005$ for the $qg$, $\overline{q}g$, and $q\overline{q}$ channels. We have studied these
channels in greater detail since they receive contributions from both the $ggb\overline{b}$ amplitudes and the four-quark amplitudes, and therefore have the most intricate 
IR structure. Results are also shown for $gg$ and $qq$ fluxes, which we have constrained to the (still stringent) level of $\epsilon \sim 0.001$, $\epsilon \sim 0.002$, or better. 
We are therefore confident that the cancellation between the unintegrated and integrated dipoles has been correctly implemented in our NLO $H+2j$ calculation.

 \begin{figure}
\begin{center}
\includegraphics[width=10cm]{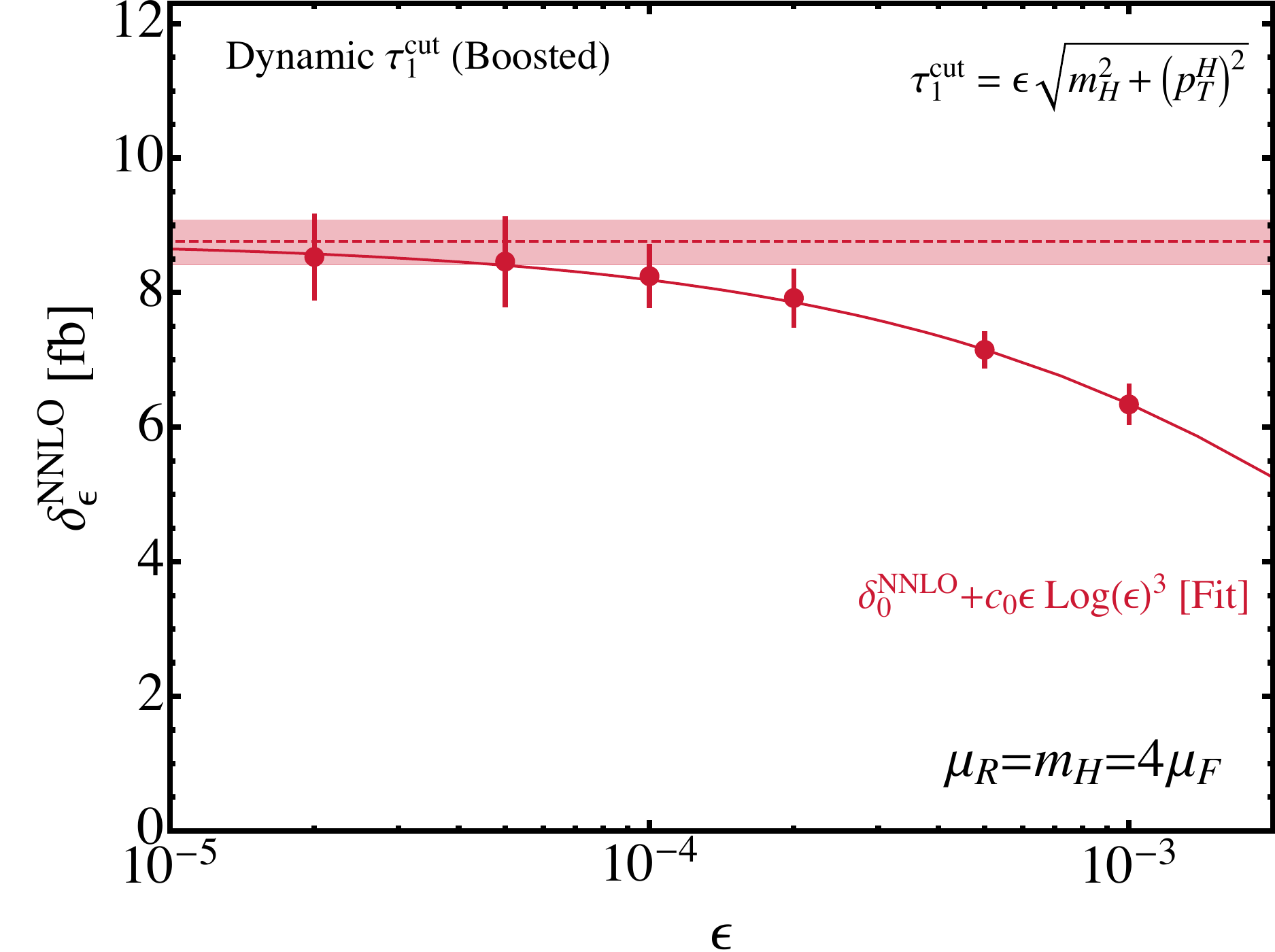}
\caption{The $\tcut1$ dependence of the NNLO coefficient for $H+j$ production at the 13 TeV LHC. Results are presented for the
 dynamic version of $\tcut1$, evaluated in the rest frame of the $H+j$ system. Also shown is a fit to the results parametrizing the residual $\epsilon$ $(\tcut1)$ 
 power corrections. The dashed line corresponds to the limit $\epsilon \rightarrow 0$ of the fit, and the shaded band represents fitting uncertainties on the asymptotic limit.   }
\label{fig:taunnloh1j}
\end{center}
\end{figure}

Finally, we arrive at the main result of this section, namely the validation of the $\tcut1$ dependence of our result for the NNLO coefficient.
We return to our previous setup used in the validation of $H+0j$ at NNLO and $H+j$ at NLO, namely we use MMHT 14 PDF sets at NNLO accuracy.  We use the canonical 
scale choices of $\mu_R = m_H = 4\,\mu_F$ and the three-loop running for the bottom quark Yukawa coupling and $\alpha_s$, such that $m^{\MS}_b(m_H)=2.793$ GeV.  Our results are presented in 
fig.~\ref{fig:taunnloh1j}. We study the dependence of the NNLO coefficient on the dynamic version of $\tcut1$, which we recall is defined as
\begin{eqnarray}
\tcut1= \epsilon \sqrt{m_H^2 + (p_T^H)^2} \, .
\end{eqnarray} 
 The results of ref.~\cite{Campbell:2019gmd} for $H+j$ in gluon fusion, and our preceding study for this process at NLO, clearly demonstrate that this choice results in the smallest power corrections, 
 particularly when the associated jet is not required to be central (as in our case). The results in fig.~\ref{fig:taunnloh1j} span the range $2 \times 10^{-5} \le \epsilon \le 1\times 10^{-3}$, which is approximately 
 equivalent to a fixed $\tcut1$ in the range $0.0025 - 0.12$ GeV  (setting $p_T^H = 30$ GeV, which corresponds to the minimum jet transverse momentum). As expected, our results are well described by the following 
 approximation for the power corrections, 
 \begin{eqnarray}
 \delta^{\rm{NNLO}}_{\epsilon} = \delta^{\rm{NNLO}}_0 + c_0\, \epsilon \log{\epsilon}^3 \, ,
 \end{eqnarray}
where $\delta^{\rm{NNLO}}_0$ represents the physical correction obtained in the limit $\epsilon \rightarrow 0$. We find 
\begin{eqnarray}
\delta_0^{\rm{NNLO}} = 8.79 \pm 0.35 \quad {\rm{fb}}
\end{eqnarray}
and, when added to the NLO cross section, we obtain
\begin{eqnarray}
\sigma_{H+j}^{\rm{NNLO}}(\mu_R = 4\mu_F = m_H) = 153.78 \pm 0.35 \, (\rm{fit}) \, {fb} \, ,
\end{eqnarray}
which means that we are able to control the remaining unknown power corrections to the level of 0.2\% on the NNLO cross section.  For the remainder of this paper we will use the boosted dynamic $\tcut1$ with 
$\epsilon$ set to $1\times 10^{-4}$. From our preceding study we can estimate that for this value the remaining power corrections should be at the level of a few percent on the NNLO coefficient, and hence 
around the per-mille level on the full physical NNLO prediction. Such a level of accuracy should be more than adequate for the phenomenology presented in the next section. 

\section{Results} 
\label{section:results}

In this section we present our results for the NNLO predictions for $H+j$ at the 13 TeV LHC. We use the  CT14 PDF sets~\cite{Dulat:2015mca}, matched to the appropriate 
order in perturbation theory (the running of $\alpha_S$ and $m_b^{\MS}(\mu_R)$ therefore occurs at the next perturbative order).  We use the same fiducial cuts as in section~\ref{section:val}, 
namely we cluster jets using the anti-$k_T$ algorithm with $R=0.4$ and require them to have $p_T^{j} > 30$ GeV and $|\eta_{j}| < 4.5$. 

\subsection{Factorization and renormalization scale dependence} 

\begin{figure}
\begin{center}
\includegraphics[width=7.45cm]{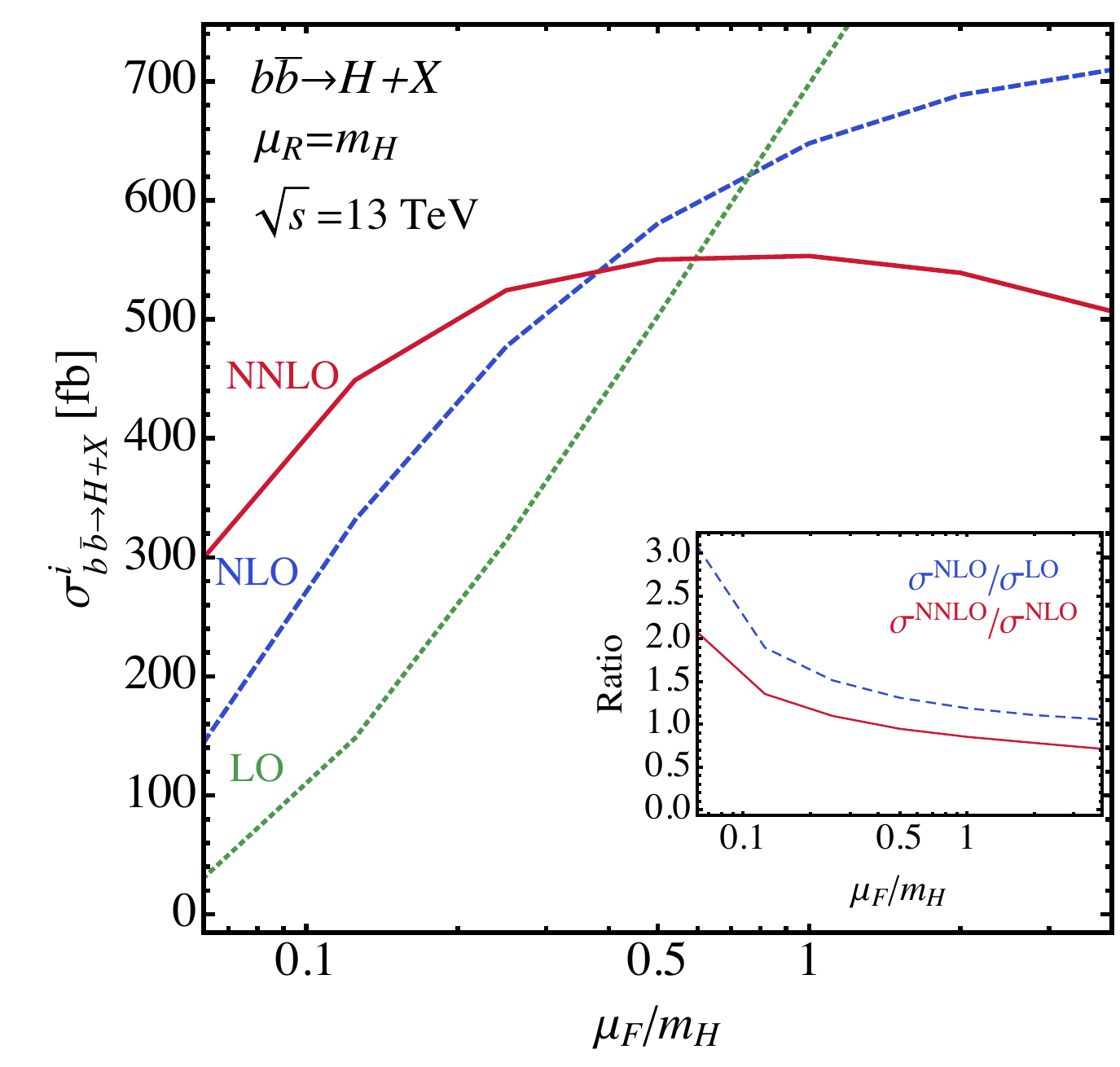}
\includegraphics[width=7.45cm]{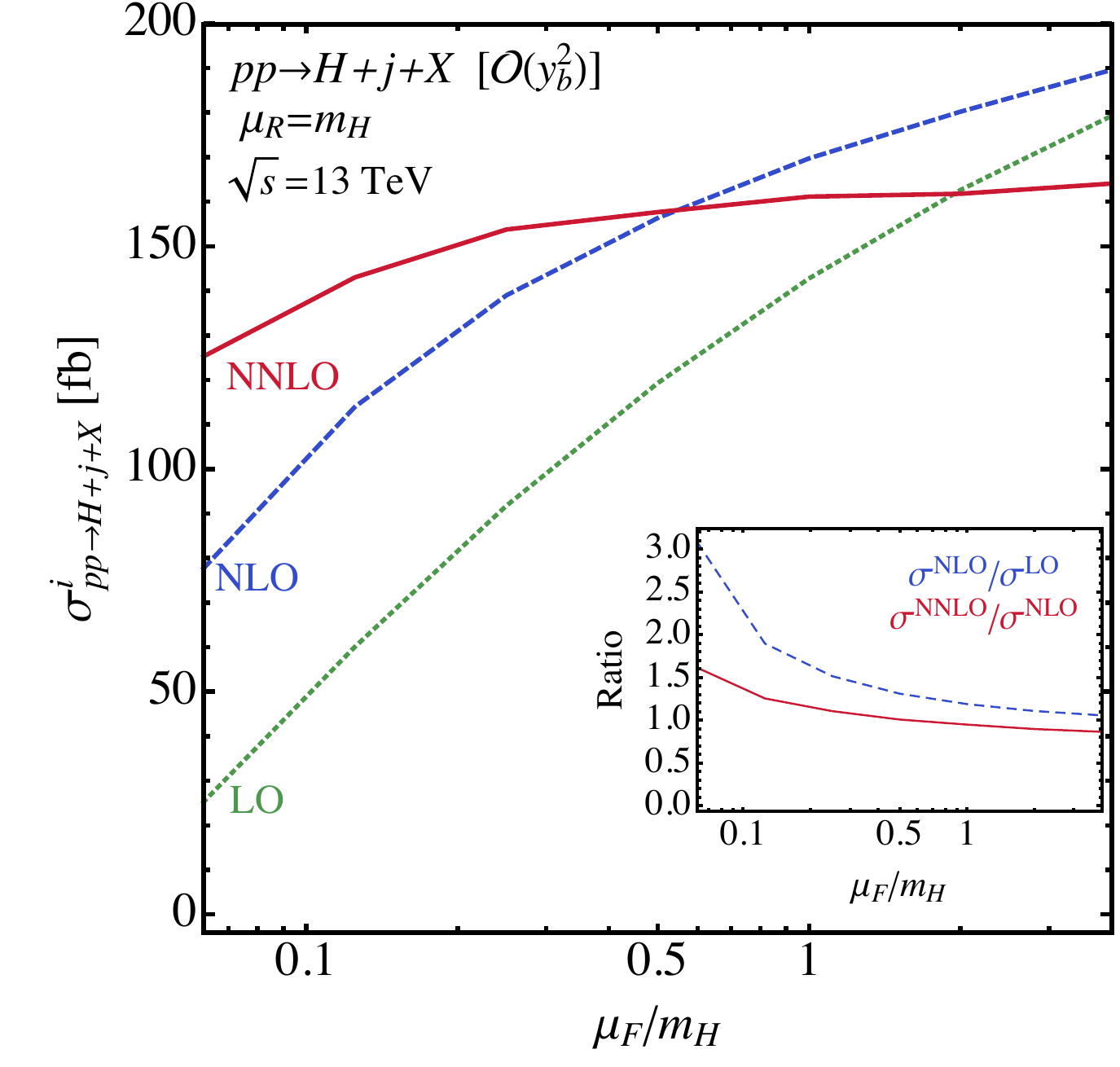}
\caption{Main figures: dependence of the LO, NLO, and NNLO cross sections on the factorization scale (left: $H+0j$, right: $H+1j$). Insets: ratio of the NNLO prediction to the NLO (red) and ratio of the NLO prediction to LO (blue, dashed).}
\label{fig:mufH0j}
\end{center}
\end{figure}

We begin by investigating the factorization and renormalization scale dependence of the total cross section for $H+1j$ at NNLO accuracy. Additionally, we also investigate 
the cross section for $H+0j$ production at the same order. Although higher-order predictions are now available~\cite{Duhr:2019kwi}, it is nevertheless interesting to compare the two predictions with and without the additional jet requirement at NNLO. 

Our results for $H+0j$ and $H+1j$ are shown in fig.~\ref{fig:mufH0j}. We set a central renormalization scale of $\mu_R=m_H$ and vary the factorization 
scale in the range $ m_H/16 \le \mu_F \le 4 m_H$.  As is well known, the factorization scale dependence for the $H+0j$ cross section is dramatic at LO and NLO, while at NNLO the behavior is somewhat improved (and even more so at N3LO~\cite{Duhr:2019kwi}). The 
maximum value around $\mu_F= m_H/3$ adds weight to the historical argument of using $\mu_F =m_H/4$ as the central scale choice in NLO predictions~\cite{Maltoni:2003pn}.

The presence of initial-state gluons and a final-state jet conspire to decrease the dependence of the $H+1j$ cross section on the factorization scale when compared to the equivalent $H+0j$ result. 
However, it is clear that the $H+j$ cross section still bears a striking dependence on the factorization scale. At LO, across the range studied the cross section increases by a factor of 7, while at NLO the increase is a factor of two. The NNLO results from our calculation lead to a substantial improvement. By including second-order corrections, the overall increase in the cross section over the range of $\mu_F$ drops to a factor of 1.36. Indeed, the vast majority of this increase occurs at lower scale choices, while at larger scales $m_H \le \mu_F \le 4 m_H$ the NNLO cross section changes only by around 6\% (compared to a change of 36\% at NLO over the same range of $\mu_F$). It is therefore clear that NNLO accuracy is mandated for a 
robust estimate of rates, free from large uncertainties induced by the unphysical dependence on the factorization scale. As the dependence on $\mu_F$ significantly drops for $\mu_F \ge m_H/4$, we will choose a central scale choice that reflects this in our subsequent predictions.

\begin{figure}
\begin{center}
\includegraphics[width=7.45cm]{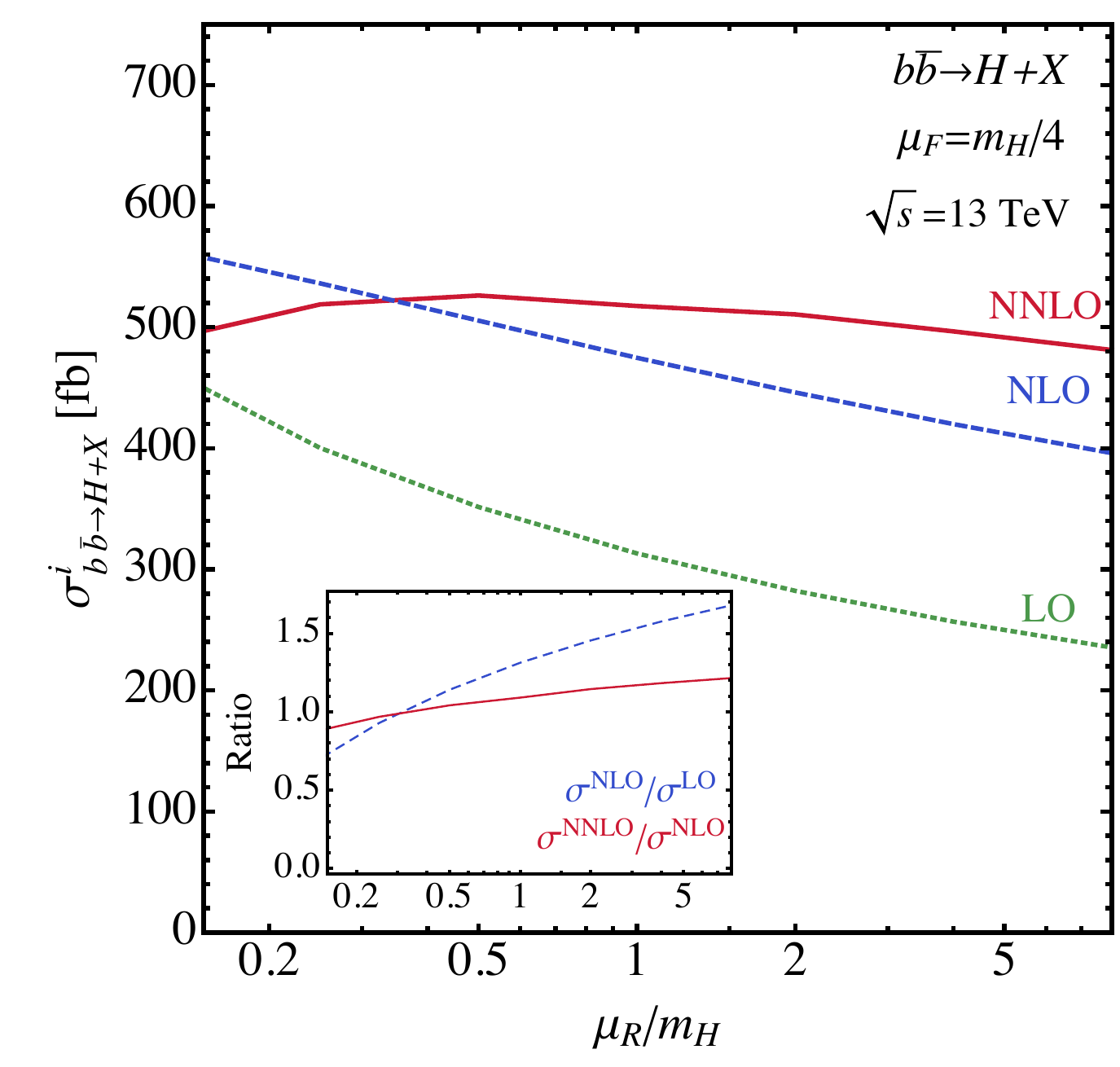}
\includegraphics[width=7.45cm]{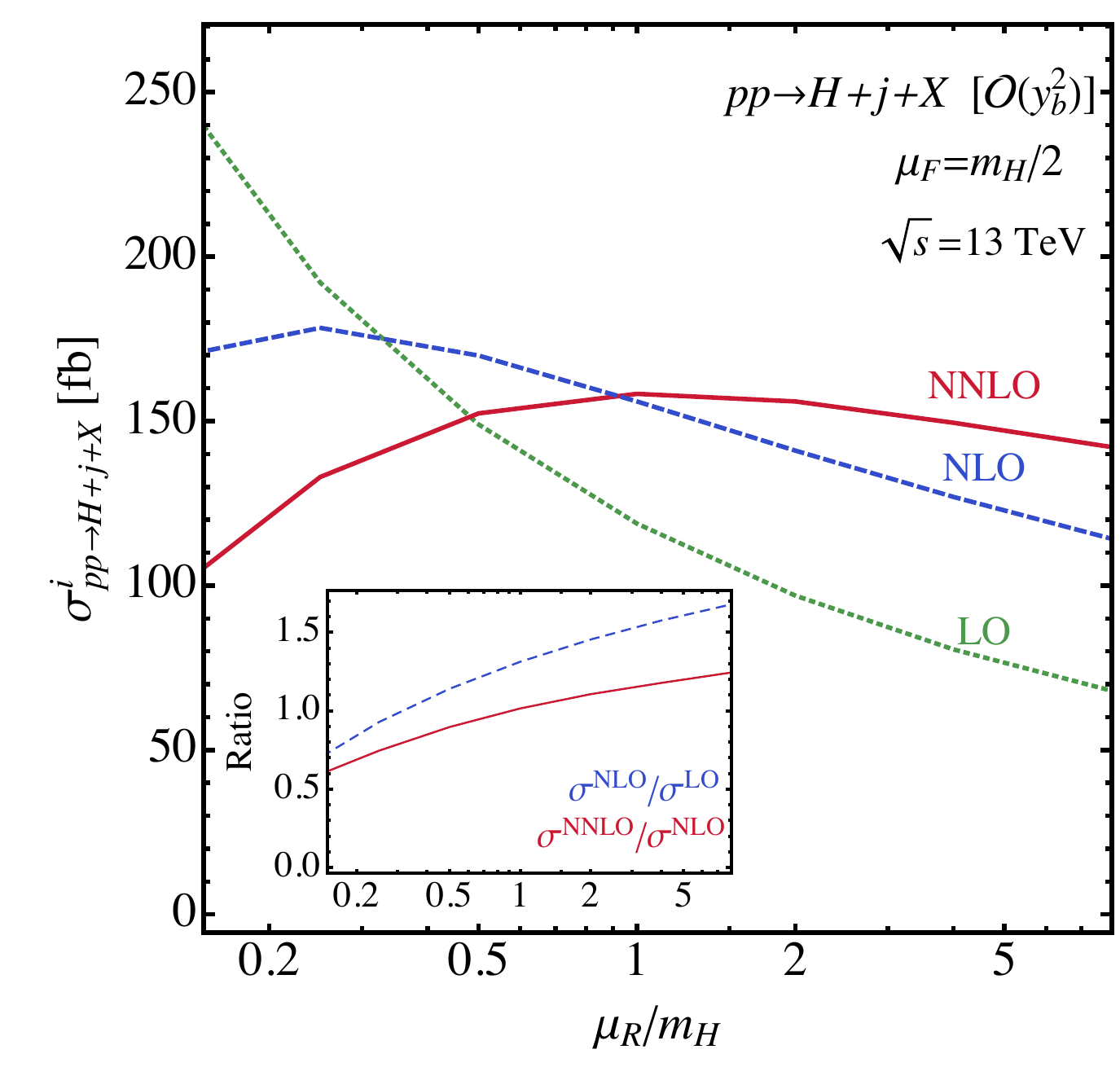}
\caption{Main figures: dependence of the LO, NLO, and NNLO cross sections on the renormalization scale (left: $H+0j$, right: $H+1j$). Insets: ratio of the NNLO prediction to the NLO (red) and ratio of the NLO prediction to LO (blue, dashed).}
\label{fig:murH0j}
\end{center}
\end{figure}

In fig.~\ref{fig:murH0j} we turn our attention to the renormalization scale dependence of the NNLO $H+0\,j$ and $H+1\,j$ cross sections. For the $b\overline{b}\rightarrow H$ process we make the customary choice $\mu_F = m_H/4$ and, motivated by the results discussed in the preceding paragraphs, we choose $\mu_F = m_H/2$ for the $H+1j$ predictions. We present the dependence of the cross sections over the range $m_H / 8 \le \mu_R \le 8 m_H$. 
For the $b\overline{b}\rightarrow H$ process, at leading order the only dependence of the cross section on $\mu_R$ arises from the evolution of $m_b^{\MS}$. Higher-order corrections induce a rather mild dependence through $\alpha_s$, and the NNLO prediction is already rather insensitive to the renormalization scale.  For the $H+j$ cross section the situation is rather different, since here the LO result depends on both $\alpha_s$ and $m_b^{\MS}$.
On the smaller end of the considered range $\mu_R < m_H/2$, the perturbation theory becomes rather unreliable, with large corrections at each subsequent order. However, in the region $m_H \le \mu_R \le 8 m_H$ the perturbation theory becomes well-behaved. There is also a significant reduction in the residual scale uncertainty at NNLO: $\sigma_{pp\rightarrow H+j}(\mu_R = m_H)/\sigma_{pp\rightarrow H+j}(\mu_R = 8 m_H)$ is $\sim 1.37$ at NLO, but reduces to $\sim 1.11$ at NNLO. 

The results of this subsection indicate that predictions obtained using central scale choices comparable to $(\mu_R,\mu_F) =  (m_H,m_H/2)$ should demonstrate convergent behavior in the perturbative expansion, with a reasonably small residual dependence for excursions from this central choice. We therefore choose these values for the differential predictions presented in the next section.

\subsection{Differential distributions}

We turn our attention to differential distributions, focusing exclusively on our new results for the $H+1j$ process. Our parameter choices are the same as those in the previous section and the central scale choice is taken to be $(\mu_R,\mu_F) =  (m_H,m_H/2)$. In order to assess the impact of the residual dependence on the renormalization and factorization scales we vary these choices using a six-point variation. Specifically, we compute our predictions with $\mu_R$ and $\mu_F$ varied by factors of two, i.e.~we compute distributions with $(\mu_R/m_H,\mu_F/(2m_H)) = (\alpha,\beta)$ where $(\alpha,\beta) \in \{(1,1),(1,2),(1,1/2),(1/2,1),(2,1),(2,2),(1/2,1/2)\}$ (without being increased and decreased in opposite directions). For each bin in the distribution the largest deviations from the central value are taken as upper and lower estimates of the scale variation. 

 \begin{figure}
\begin{center}
\includegraphics[width=7.49cm]{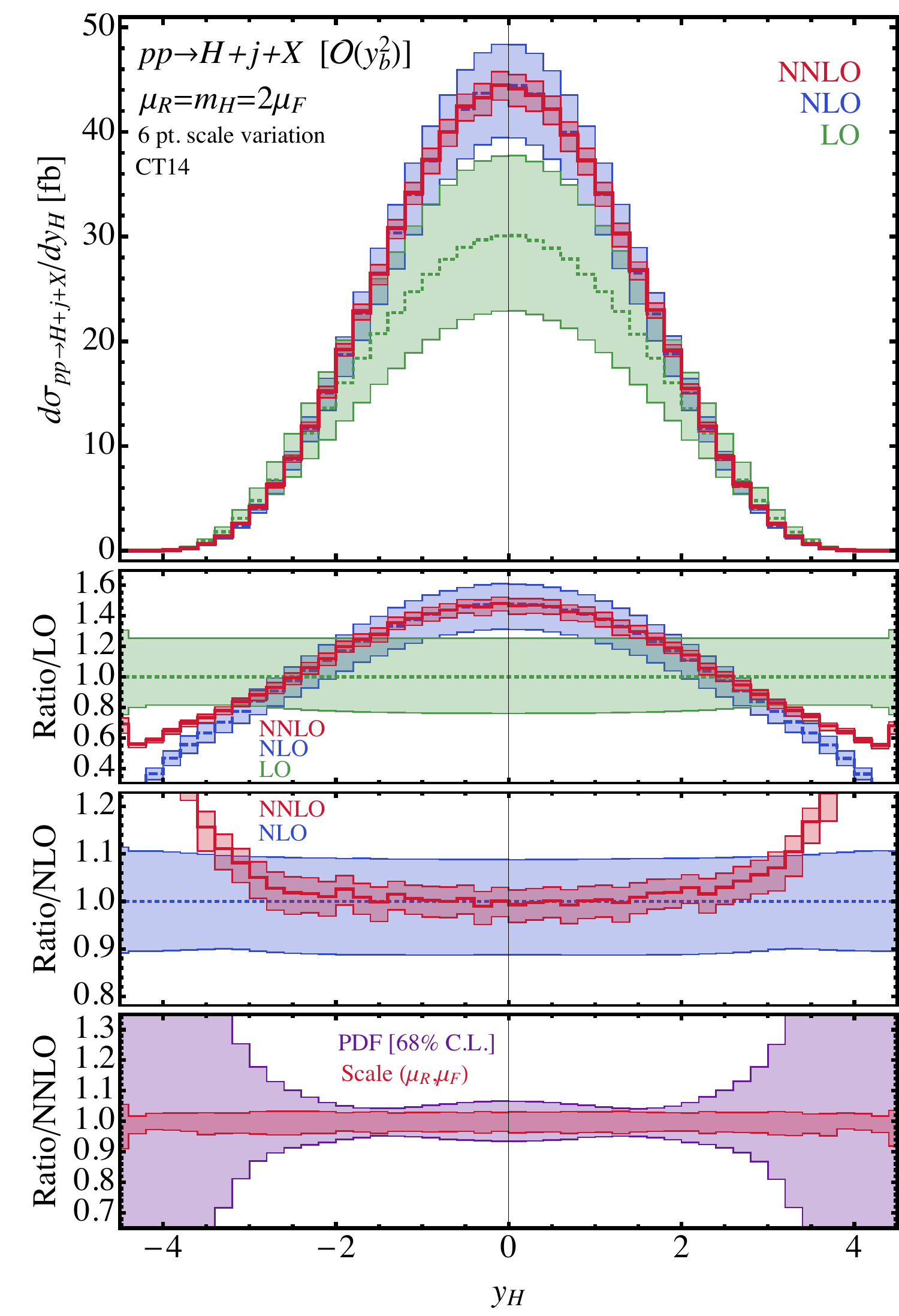}
\includegraphics[width=7.49cm]{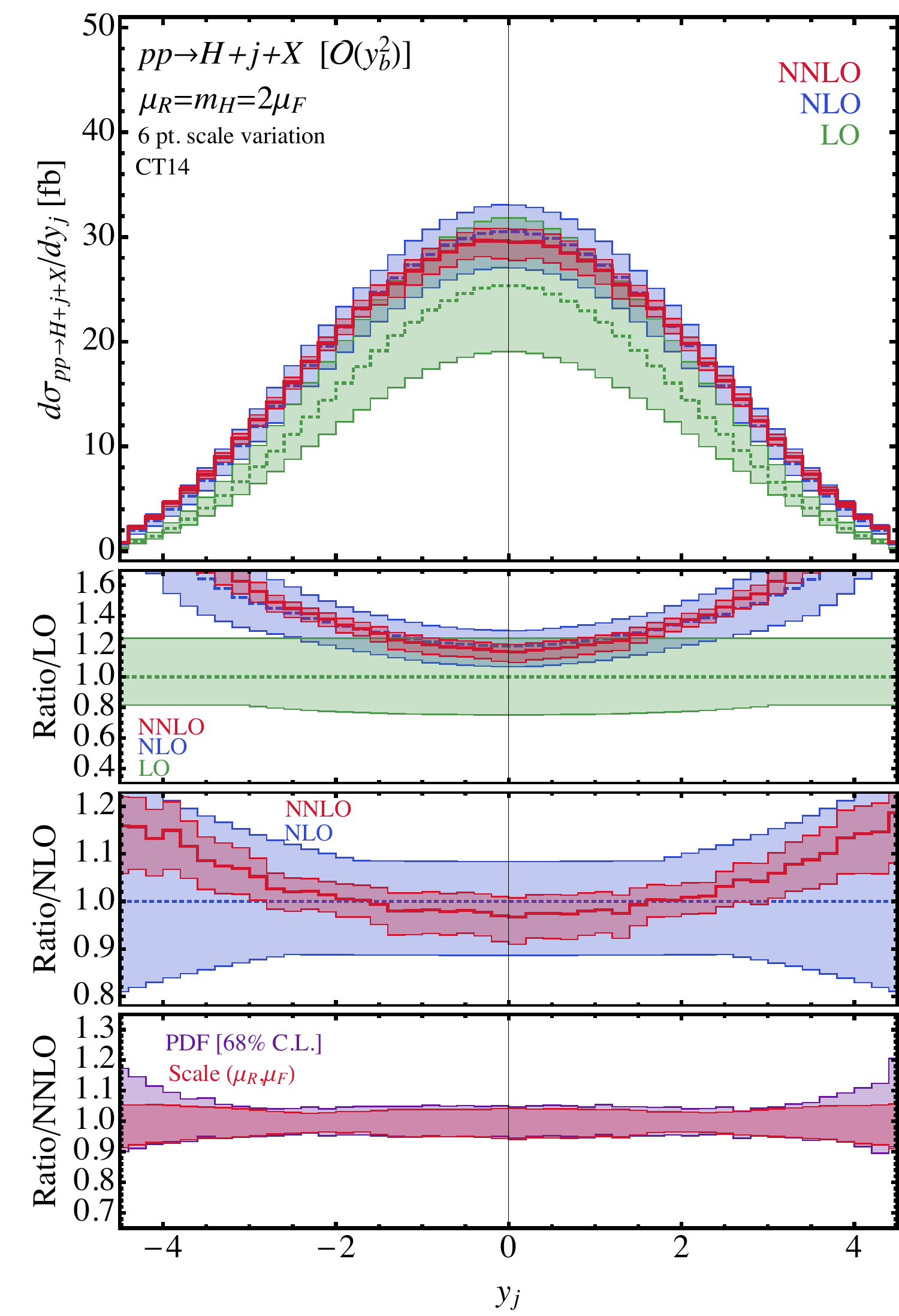}
\caption{Distributions of the rapidity of the Higgs boson ($y_H$, left) and leading jet ($y_j$, right) at LO, NLO, and NNLO accuracy. The top panel shows the distribution, while the lower panels present ratios to the LO, NLO, and NNLO central predictions. 
In addition to the six-point scale variation, the lower panel presents estimates of PDF uncertainties at 68\% C.L. (the other panels only include the scale variation).}
\label{fig:h1jyrap}
\end{center}
\end{figure}

Fig.~\ref{fig:h1jyrap} shows the results for the rapidity distribution of the Higgs boson ($y_H$) and leading jet ($y_j$) at LO, NLO, and NNLO accuracy. By comparing the two distributions it is clear that the Higgs boson is produced with a more central distribution than the jet, which has a broader distribution (and hence more forward jets). This can be traced back to the underlying kinematics, since the Higgs boson is a scalar particle and therefore is produced more isotropically, while the leading jet favours the collinear (forward) region in which the quark-gluon splitting is enhanced. The pattern of higher-order corrections is broadly similar for both distributions, with a significant shape change from LO to NLO and a much smaller change from NLO to NNLO. We observe that the scale variation decreases from around $\pm 10$\% at NLO to around $\pm 4\%$ at NNLO. In the central region $|y_X| < 2.5$ ($X=H,j$), the corrections are relatively flat, whereas in the larger rapidity regions they become more sizable. We note that some care should be taken in this region, since it could be prone to larger power corrections in the $N$-jettiness slicing method. However, we estimate that remaining power corrections enter at around the percent level in the tails of the two distributions, which should not substantially change the interpretation of the plots. In the lower panel we additionally include an estimate of the uncertainties due to the PDF extractions, obtained at 68\% C.L. using LHAPDF~\cite{Buckley:2014ana}. We see that at NNLO the uncertainties from the PDFs and the six-point scale variation are of the same order ($\sim 5\%$). For very forward Higgs bosons the PDF uncertainties become very large, but there is very little cross section in this region. 

 \begin{figure}
\begin{center}
\includegraphics[width=7.49cm]{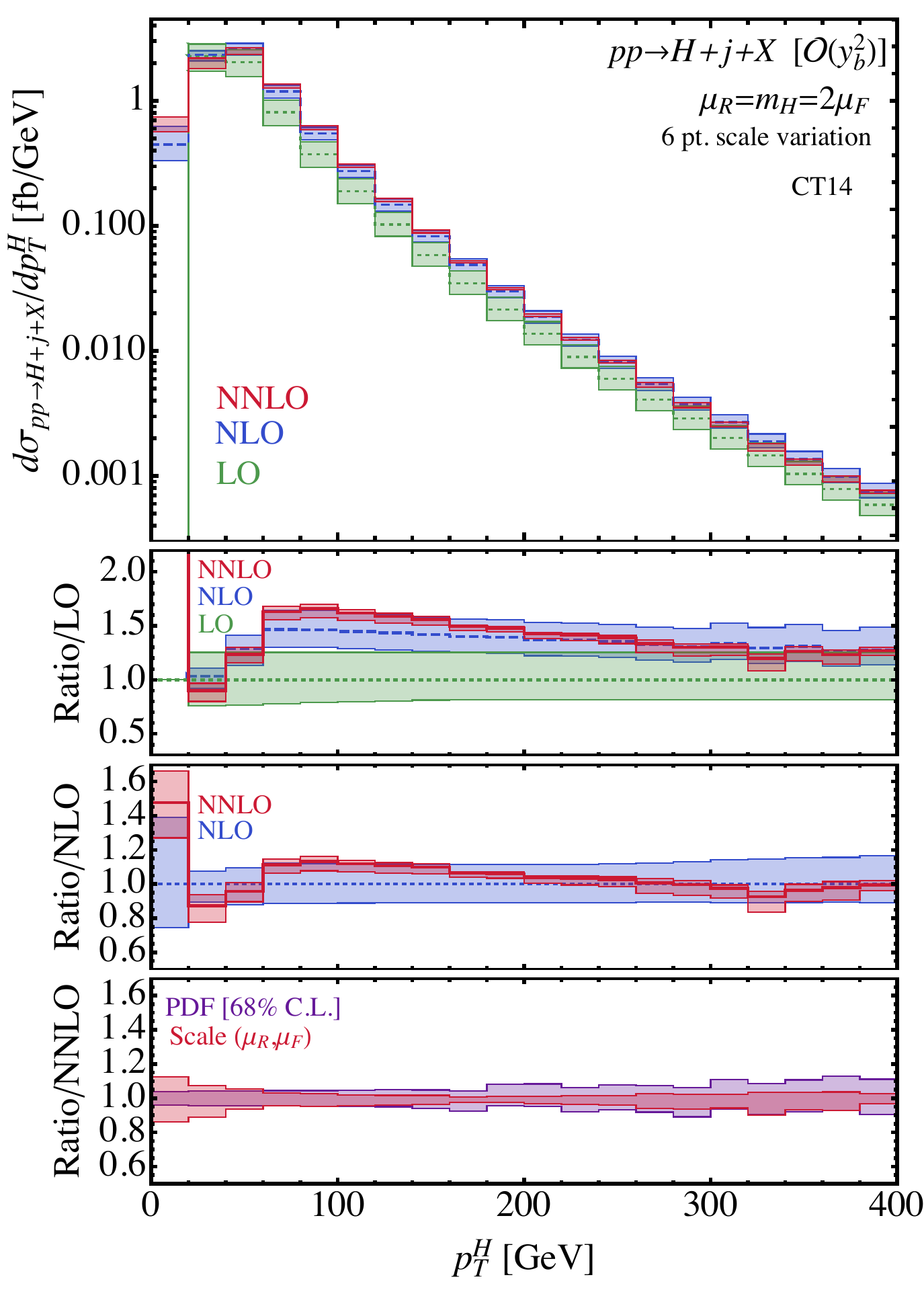}
\includegraphics[width=7.49cm]{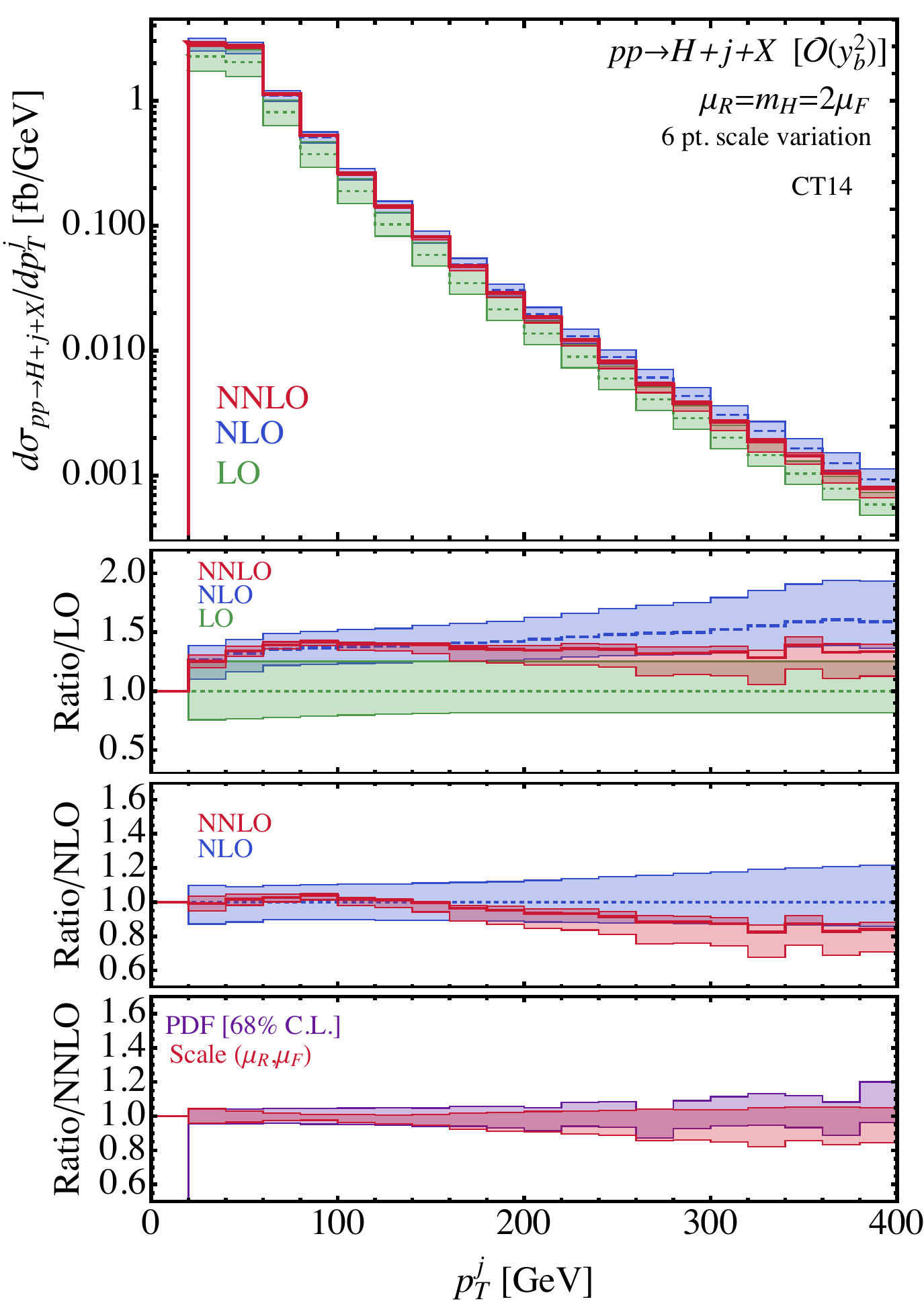}
\caption{Distributions of the transverse momentum of the Higgs boson ($p_T^H$, left) and leading jet ($p_T^j$, right) at LO, NLO, and NNLO accuracy. The top panel shows the distribution, while the lower panels present ratios to the LO, NLO, and NNLO central predictions. 
In addition to the six-point scale variation, the lower panel presents estimates of PDF uncertainties at 68\% C.L. (the other panels only include the scale variation).}
\label{fig:h1jpt}
\end{center}
\end{figure}

Fig.~\ref{fig:h1jpt} presents the transverse momentum distribution of the Higgs boson and the leading jet. For the transverse momentum of the Higgs boson the 
softest bin $p_T^H < 30$ GeV corresponds to an observable one perturbative order lower than the rest of the calculation (since there exists no $2\rightarrow 2$ underlying topology) and this is reflected in the 
larger NNLO/NLO ratio and overall scale variation of this bin. Focusing on the change from NLO to NNLO, we see that the ratios are rather 
dynamic (while remaining within the uncertainty band of the NLO calculations), especially for the transverse momentum of the Higgs boson. In the region $ p_T^H < m_H/2$ there is a decrease in the prediction of around $5-10\%$, while in the region around $p_T^H \sim m_H$ the prediction increases at NNLO by around $10-15\%$ before asymptotically approaching a smaller ratio (around $5\%$) at larger transverse momentum.  The leading-jet transverse momentum distribution is different: the main impact of the NNLO corrections is to soften the spectrum, especially at high $p_T$. Finally, for both distributions the PDF uncertainties are again comparable in size to those obtained using a six-point scale variation.

\section{Conclusions} 

\label{section:conc}

We have presented a NNLO calculation of the bottom-induced contributions to $pp\rightarrow H+j$ in the 5FS. Our results have been implemented into MCFM and, as a useful by-product and cross-check of our setup, we have also implemented Higgs boson production through bottom-quark fusion $b\overline{b} \rightarrow H$ at the same order.
Analytic results for the various Higgs-plus-parton amplitudes needed in this paper have been obtained from our previous study of the Higgs boson decay to three partons, crossed to the appropriate LHC kinematic configurations. 
We have performed many cross-checks of our calculation. At the analytic level we have checked the collinear and soft factorization properties of our two-loop amplitudes, and have carried out numerical checks of our $H+4,5$ parton amplitudes. For the $H+2j$ NLO computation we have done extensive testing on the exact cancellation of the integrated and unintegrated dipole subtractions.  
We have compared our results for $b\overline{b} \rightarrow H$ to the public code {\tt{Sushi}}, finding excellent agreement. 

Higgs plus jet is fast becoming a standard candle process to study the Higgs boson at the LHC. While the bottom-initiated contributions remain a small piece of the total cross section, their study is motivated by the strong desire to constrain the bottom Yukawa coupling wherever possible. Unfortunately, bottom-induced processes in the 5FS have a strong dependence on the factorization scale, making a precision computation of these cross sections challenging. The results of this paper show a dramatic stabilization of the cross section at NNLO, particularly with regard to the overall dependence on the factorization scale. 

While our results focused on the process $pp\rightarrow H+j$, there are several interesting extensions of this study to pursue. Firstly, in order to target the bottom Yukawa interaction, experimental analyses would need to impose $b$-tagging requirements. Therefore, adjusting our computation to target $pp\rightarrow H+b$ at NNLO is a logical next step. In order to do a meaningful phenomenological study for the LHC, the bottom-induced contributions should be compared to the dominant production mechanism through gluon fusion (with a $b$-tagging requirement applied). Dealing with the interference term in a sensible limit in the 5FS at this order is also interesting for a full phenomenology study. The inclusion of the decay of the Higgs boson is also a must in order to adequately describe fiducial volumes, for which a particularly interesting example is when the Higgs 
boson decays to bottom quarks (which must also be matched to NNLO accuracy). With these alterations in hand, it would also be interesting to study Higgs-plus-charm production. Finally, in addition to the SM Higgs boson, the calculation described here could also be used in BSM extensions. For instance, the Higgs-bottom quark coupling could be modified as in the MSSM or SMEFT, or the mass of the scalar particle itself 
could be increased, e.g.~in dark matter searches where the scalar particle acts as a mediator of putative dark forces (with potential missing energy decays). We leave these exciting and detailed studies to future work.

\acknowledgments

We thank Uli Schubert for useful discussions. We are particularly grateful to John Campbell for 
many essential discussions regarding the results of ref.~\cite{Campbell:2019gmd}.
The authors are supported by a National Science Foundation CAREER award number PHY-1652066.
Support provided by the Center for Computational Research at the University at Buffalo.

\appendix

\bibliographystyle{JHEP}

\bibliography{Hbb}

\providecommand{\href}[2]{#2}\begingroup\raggedright\begin{thebibliography}{10}

\bibitem{Aad:2012tfa}
{\scshape ATLAS} collaboration, \emph{{Observation of a new particle in the
  search for the Standard Model Higgs boson with the ATLAS detector at the
  LHC}}, \href{https://doi.org/10.1016/j.physletb.2012.08.020}{\emph{Phys.
  Lett.} {\bfseries B716} (2012) 1}
  [\href{https://arxiv.org/abs/1207.7214}{{\ttfamily 1207.7214}}].

\bibitem{Chatrchyan:2012xdj}
{\scshape CMS} collaboration, \emph{{Observation of a new boson at a mass of
  125 GeV with the CMS experiment at the LHC}},
  \href{https://doi.org/10.1016/j.physletb.2012.08.021}{\emph{Phys. Lett.}
  {\bfseries B716} (2012) 30}
  [\href{https://arxiv.org/abs/1207.7235}{{\ttfamily 1207.7235}}].

\bibitem{Aad:2014xva}
{\scshape ATLAS} collaboration, \emph{{Search for the Standard Model Higgs
  boson decay to $\mu^{+}\mu^{-}$ with the ATLAS detector}},
  \href{https://doi.org/10.1016/j.physletb.2014.09.008}{\emph{Phys. Lett. B}
  {\bfseries 738} (2014) 68} [\href{https://arxiv.org/abs/1406.7663}{{\ttfamily
  1406.7663}}].

\bibitem{Khachatryan:2014jba}
{\scshape CMS} collaboration, \emph{{Precise determination of the mass of the
  Higgs boson and tests of compatibility of its couplings with the standard
  model predictions using proton collisions at 7 and 8 $\,\text {TeV}$}},
  \href{https://doi.org/10.1140/epjc/s10052-015-3351-7}{\emph{Eur. Phys. J. C}
  {\bfseries 75} (2015) 212} [\href{https://arxiv.org/abs/1412.8662}{{\ttfamily
  1412.8662}}].

\bibitem{Aad:2019mbh}
{\scshape ATLAS} collaboration, \emph{{Combined measurements of Higgs boson
  production and decay using up to $80$ fb$^{-1}$ of proton-proton collision
  data at $\sqrt{s}=$ 13 TeV collected with the ATLAS experiment}},
  \href{https://doi.org/10.1103/PhysRevD.101.012002}{\emph{Phys. Rev. D}
  {\bfseries 101} (2020) 012002}
  [\href{https://arxiv.org/abs/1909.02845}{{\ttfamily 1909.02845}}].

\bibitem{Sirunyan:2018koj}
{\scshape CMS} collaboration, \emph{{Combined measurements of Higgs boson
  couplings in proton\textendash{}proton collisions at $\sqrt{s}=13\,\text
  {Te}\text {V} $}},
  \href{https://doi.org/10.1140/epjc/s10052-019-6909-y}{\emph{Eur. Phys. J. C}
  {\bfseries 79} (2019) 421}
  [\href{https://arxiv.org/abs/1809.10733}{{\ttfamily 1809.10733}}].

\bibitem{Abada:2019lih}
{\scshape FCC} collaboration, \emph{{FCC Physics Opportunities}: {Future
  Circular Collider Conceptual Design Report Volume 1}},
  \href{https://doi.org/10.1140/epjc/s10052-019-6904-3}{\emph{Eur. Phys. J. C}
  {\bfseries 79} (2019) 474}.

\bibitem{Contino:2016spe}
R.~Contino et~al., \emph{{Physics at a 100 TeV pp collider: Higgs and EW
  symmetry breaking studies}},
  \href{https://doi.org/10.23731/CYRM-2017-003.255}{\emph{CERN Yellow Rep.}
  (2017) 255} [\href{https://arxiv.org/abs/1606.09408}{{\ttfamily
  1606.09408}}].

\bibitem{Cohen:2017rsk}
J.~Cohen, S.~Bar-Shalom, G.~Eilam and A.~Soni, \emph{{Light-quarks Yukawa
  couplings and new physics in exclusive high- $p_T$ Higgs boson+jet and Higgs
  boson + b -jet events}},
  \href{https://doi.org/10.1103/PhysRevD.97.055014}{\emph{Phys. Rev. D}
  {\bfseries 97} (2018) 055014}
  [\href{https://arxiv.org/abs/1705.09295}{{\ttfamily 1705.09295}}].

\bibitem{Arcadi:2019lka}
G.~Arcadi, A.~Djouadi and M.~Raidal, \emph{{Dark Matter through the Higgs
  portal}}, \href{https://doi.org/10.1016/j.physrep.2019.11.003}{\emph{Phys.
  Rept.} {\bfseries 842} (2020) 1}
  [\href{https://arxiv.org/abs/1903.03616}{{\ttfamily 1903.03616}}].

\bibitem{Dawson:2011pe}
S.~Dawson, C.~Jackson and P.~Jaiswal, \emph{{SUSY QCD Corrections to Higgs-b
  Production : Is the $\Delta_b$ Approximation Accurate?}},
  \href{https://doi.org/10.1103/PhysRevD.83.115007}{\emph{Phys. Rev. D}
  {\bfseries 83} (2011) 115007}
  [\href{https://arxiv.org/abs/1104.1631}{{\ttfamily 1104.1631}}].

\bibitem{Dawson:2007ur}
S.~Dawson and C.~Jackson, \emph{{SUSY QCD Corrections to Associated
  Higgs-bottom Quark Production}},
  \href{https://doi.org/10.1103/PhysRevD.77.015019}{\emph{Phys. Rev. D}
  {\bfseries 77} (2008) 015019}
  [\href{https://arxiv.org/abs/0709.4519}{{\ttfamily 0709.4519}}].

\bibitem{Dicus:1998hs}
D.~Dicus, T.~Stelzer, Z.~Sullivan and S.~Willenbrock, \emph{{Higgs boson
  production in association with bottom quarks at next-to-leading order}},
  \href{https://doi.org/10.1103/PhysRevD.59.094016}{\emph{Phys. Rev. D}
  {\bfseries 59} (1999) 094016}
  [\href{https://arxiv.org/abs/hep-ph/9811492}{{\ttfamily hep-ph/9811492}}].

\bibitem{Balazs:1998sb}
C.~Balazs, H.-J. He and C.~P. Yuan, \emph{{QCD corrections to scalar production
  via heavy quark fusion at hadron colliders}},
  \href{https://doi.org/10.1103/PhysRevD.60.114001}{\emph{Phys. Rev. D}
  {\bfseries 60} (1999) 114001}
  [\href{https://arxiv.org/abs/hep-ph/9812263}{{\ttfamily hep-ph/9812263}}].

\bibitem{Campbell:2002zm}
J.~M. Campbell, R.~Ellis, F.~Maltoni and S.~Willenbrock, \emph{{Higgs-Boson
  Production in Association with a Single Bottom Quark}},
  \href{https://doi.org/10.1103/PhysRevD.67.095002}{\emph{Phys. Rev. D}
  {\bfseries 67} (2003) 095002}
  [\href{https://arxiv.org/abs/hep-ph/0204093}{{\ttfamily hep-ph/0204093}}].

\bibitem{Maltoni:2003pn}
F.~Maltoni, Z.~Sullivan and S.~Willenbrock, \emph{{Higgs-Boson Production via
  Bottom-Quark Fusion}},
  \href{https://doi.org/10.1103/PhysRevD.67.093005}{\emph{Phys. Rev. D}
  {\bfseries 67} (2003) 093005}
  [\href{https://arxiv.org/abs/hep-ph/0301033}{{\ttfamily hep-ph/0301033}}].

\bibitem{Dawson:2003kb}
S.~Dawson, C.~Jackson, L.~Reina and D.~Wackeroth, \emph{{Exclusive Higgs boson
  production with bottom quarks at hadron colliders}},
  \href{https://doi.org/10.1103/PhysRevD.69.074027}{\emph{Phys. Rev. D}
  {\bfseries 69} (2004) 074027}
  [\href{https://arxiv.org/abs/hep-ph/0311067}{{\ttfamily hep-ph/0311067}}].

\bibitem{Dawson:2004sh}
S.~Dawson, C.~Jackson, L.~Reina and D.~Wackeroth, \emph{{Higgs boson production
  with one bottom quark jet at hadron colliders}},
  \href{https://doi.org/10.1103/PhysRevLett.94.031802}{\emph{Phys. Rev. Lett.}
  {\bfseries 94} (2005) 031802}
  [\href{https://arxiv.org/abs/hep-ph/0408077}{{\ttfamily hep-ph/0408077}}].

\bibitem{Harlander:2010cz}
R.~V. Harlander, K.~J. Ozeren and M.~Wiesemann, \emph{{Higgs plus jet
  production in bottom quark annihilation at next-to-leading order}},
  \href{https://doi.org/10.1016/j.physletb.2010.08.038}{\emph{Phys. Lett. B}
  {\bfseries 693} (2010) 269}
  [\href{https://arxiv.org/abs/1007.5411}{{\ttfamily 1007.5411}}].

\bibitem{Harlander:2011aa}
R.~Harlander, M.~Kramer and M.~Schumacher, \emph{{Bottom-quark associated
  Higgs-boson production: reconciling the four- and five-flavour scheme
  approach}},  \href{https://arxiv.org/abs/1112.3478}{{\ttfamily 1112.3478}}.

\bibitem{Buehler:2012cu}
S.~B\"uhler, F.~Herzog, A.~Lazopoulos and R.~M\"uller, \emph{{The fully
  differential hadronic production of a Higgs boson via bottom quark fusion at
  NNLO}}, \href{https://doi.org/10.1007/JHEP07(2012)115}{\emph{JHEP} {\bfseries
  07} (2012) 115} [\href{https://arxiv.org/abs/1204.4415}{{\ttfamily
  1204.4415}}].

\bibitem{Harlander:2012pb}
R.~V. Harlander, S.~Liebler and H.~Mantler, \emph{{SusHi: A program for the
  calculation of Higgs production in gluon fusion and bottom-quark annihilation
  in the Standard Model and the MSSM}},
  \href{https://doi.org/10.1016/j.cpc.2013.02.006}{\emph{Comput. Phys. Commun.}
  {\bfseries 184} (2013) 1605}
  [\href{https://arxiv.org/abs/1212.3249}{{\ttfamily 1212.3249}}].

\bibitem{Maltoni:2012pa}
F.~Maltoni, G.~Ridolfi and M.~Ubiali, \emph{{b-initiated processes at the LHC:
  a reappraisal}}, \href{https://doi.org/10.1007/JHEP04(2013)095}{\emph{JHEP}
  {\bfseries 07} (2012) 022} [\href{https://arxiv.org/abs/1203.6393}{{\ttfamily
  1203.6393}}].

\bibitem{Wiesemann:2014ioa}
M.~Wiesemann, R.~Frederix, S.~Frixione, V.~Hirschi, F.~Maltoni and
  P.~Torrielli, \emph{{Higgs production in association with bottom quarks}},
  \href{https://doi.org/10.1007/JHEP02(2015)132}{\emph{JHEP} {\bfseries 02}
  (2015) 132} [\href{https://arxiv.org/abs/1409.5301}{{\ttfamily 1409.5301}}].

\bibitem{Bonvini:2015pxa}
M.~Bonvini, A.~S. Papanastasiou and F.~J. Tackmann, \emph{{Resummation and
  matching of b-quark mass effects in $ b\overline{b}H $ production}},
  \href{https://doi.org/10.1007/JHEP11(2015)196}{\emph{JHEP} {\bfseries 11}
  (2015) 196} [\href{https://arxiv.org/abs/1508.03288}{{\ttfamily
  1508.03288}}].

\bibitem{Jager:2015hka}
B.~Jager, L.~Reina and D.~Wackeroth, \emph{{Higgs boson production in
  association with b jets in the POWHEG BOX}},
  \href{https://doi.org/10.1103/PhysRevD.93.014030}{\emph{Phys. Rev. D}
  {\bfseries 93} (2016) 014030}
  [\href{https://arxiv.org/abs/1509.05843}{{\ttfamily 1509.05843}}].

\bibitem{Forte:2016sja}
S.~Forte, D.~Napoletano and M.~Ubiali, \emph{{Higgs production in bottom-quark
  fusion: matching beyond leading order}},
  \href{https://doi.org/10.1016/j.physletb.2016.10.040}{\emph{Phys. Lett. B}
  {\bfseries 763} (2016) 190}
  [\href{https://arxiv.org/abs/1607.00389}{{\ttfamily 1607.00389}}].

\bibitem{Krauss:2016orf}
F.~Krauss, D.~Napoletano and S.~Schumann, \emph{{Simulating $b$-associated
  production of $Z$ and Higgs bosons with the SHERPA event generator}},
  \href{https://doi.org/10.1103/PhysRevD.95.036012}{\emph{Phys. Rev. D}
  {\bfseries 95} (2017) 036012}
  [\href{https://arxiv.org/abs/1612.04640}{{\ttfamily 1612.04640}}].

\bibitem{Deutschmann:2018avk}
N.~Deutschmann, F.~Maltoni, M.~Wiesemann and M.~Zaro, \emph{{Top-Yukawa
  contributions to bbH production at the LHC}},
  \href{https://doi.org/10.1007/JHEP07(2019)054}{\emph{JHEP} {\bfseries 07}
  (2019) 054} [\href{https://arxiv.org/abs/1808.01660}{{\ttfamily
  1808.01660}}].

\bibitem{H:2019nsw}
A.~Ajjath, P.~Banerjee, A.~Chakraborty, P.~K. Dhani, P.~Mukherjee, N.~Rana
  et~al., \emph{{NNLO QCD$\oplus$QED corrections to Higgs production in bottom
  quark annihilation}},
  \href{https://doi.org/10.1103/PhysRevD.100.114016}{\emph{Phys. Rev. D}
  {\bfseries 100} (2019) 114016}
  [\href{https://arxiv.org/abs/1906.09028}{{\ttfamily 1906.09028}}].

\bibitem{Pagani:2020rsg}
D.~Pagani, H.-S. Shao and M.~Zaro, \emph{{RIP $ Hb\overline{b} $: how other
  Higgs production modes conspire to kill a rare signal at the LHC}},
  \href{https://doi.org/10.1007/JHEP11(2020)036}{\emph{JHEP} {\bfseries 11}
  (2020) 036} [\href{https://arxiv.org/abs/2005.10277}{{\ttfamily
  2005.10277}}].

\bibitem{Grojean:2020ech}
C.~Grojean, A.~Paul and Z.~Qian, \emph{{Resurrecting $b\bar{b}h$ with kinematic
  shapes}},  \href{https://arxiv.org/abs/2011.13945}{{\ttfamily 2011.13945}}.

\bibitem{Duhr:2019kwi}
C.~Duhr, F.~Dulat and B.~Mistlberger, \emph{{Higgs Boson Production in
  Bottom-Quark Fusion to Third Order in the Strong Coupling}},
  \href{https://doi.org/10.1103/PhysRevLett.125.051804}{\emph{Phys. Rev. Lett.}
  {\bfseries 125} (2020) 051804}
  [\href{https://arxiv.org/abs/1904.09990}{{\ttfamily 1904.09990}}].

\bibitem{Duhr:2020kzd}
C.~Duhr, F.~Dulat, V.~Hirschi and B.~Mistlberger, \emph{{Higgs production in
  bottom quark fusion: matching the 4- and 5-flavour schemes to third order in
  the strong coupling}},
  \href{https://doi.org/10.1007/JHEP08(2020)017}{\emph{JHEP} {\bfseries 08}
  (2020) 017} [\href{https://arxiv.org/abs/2004.04752}{{\ttfamily
  2004.04752}}].

\bibitem{Plehn:2002vy}
T.~Plehn, \emph{{Charged Higgs boson production in bottom gluon fusion}},
  \href{https://doi.org/10.1103/PhysRevD.67.014018}{\emph{Phys. Rev. D}
  {\bfseries 67} (2003) 014018}
  [\href{https://arxiv.org/abs/hep-ph/0206121}{{\ttfamily hep-ph/0206121}}].

\bibitem{Chetyrkin:1996sr}
K.~G. Chetyrkin, \emph{{Correlator of the quark scalar currents and
  {$\Gamma_{tot}$} (H ---> hadrons) at {$\mathcal{O}$}({$\alpha_{s}^{3}$}) in
  pQCD}}, \href{https://doi.org/10.1016/S0370-2693(96)01368-8}{\emph{Phys.
  Lett.} {\bfseries B390} (1997) 309}
  [\href{https://arxiv.org/abs/hep-ph/9608318}{{\ttfamily hep-ph/9608318}}].

\bibitem{Anastasiou:2011qx}
C.~Anastasiou, F.~Herzog and A.~Lazopoulos, \emph{{The fully differential decay
  rate of a Higgs boson to bottom-quarks at NNLO in QCD}},
  \href{https://doi.org/10.1007/JHEP03(2012)035}{\emph{JHEP} {\bfseries 03}
  (2012) 035} [\href{https://arxiv.org/abs/1110.2368}{{\ttfamily 1110.2368}}].

\bibitem{DelDuca:2015zqa}
V.~Del~Duca, C.~Duhr, G.~Somogyi, F.~Tramontano and Z.~Tr\'ocs\'anyi,
  \emph{{Higgs boson decay into b-quarks at NNLO accuracy}},
  \href{https://doi.org/10.1007/JHEP04(2015)036}{\emph{JHEP} {\bfseries 04}
  (2015) 036} [\href{https://arxiv.org/abs/1501.07226}{{\ttfamily
  1501.07226}}].

\bibitem{Baikov:2005rw}
P.~A. Baikov, K.~G. Chetyrkin and J.~H. Kuhn, \emph{{Scalar correlator at
  {$\mathcal{O}$}({$\alpha_{s}^{4}$}), Higgs decay into b-quarks and bounds on
  the light quark masses}},
  \href{https://doi.org/10.1103/PhysRevLett.96.012003}{\emph{Phys. Rev. Lett.}
  {\bfseries 96} (2006) 012003}
  [\href{https://arxiv.org/abs/hep-ph/0511063}{{\ttfamily hep-ph/0511063}}].

\bibitem{Mondini:2019gid}
R.~Mondini, M.~Schiavi and C.~Williams, \emph{{N$^{3}$LO predictions for the
  decay of the Higgs boson to bottom quarks}},
  \href{https://doi.org/10.1007/JHEP06(2019)079}{\emph{JHEP} {\bfseries 06}
  (2019) 079} [\href{https://arxiv.org/abs/1904.08960}{{\ttfamily
  1904.08960}}].

\bibitem{Bernreuther:2018ynm}
W.~Bernreuther, L.~Chen and Z.-G. Si, \emph{{Differential decay rates of
  CP-even and CP-odd Higgs bosons to top and bottom quarks at NNLO QCD}},
  \href{https://doi.org/10.1007/JHEP07(2018)159}{\emph{JHEP} {\bfseries 07}
  (2018) 159} [\href{https://arxiv.org/abs/1805.06658}{{\ttfamily
  1805.06658}}].

\bibitem{Behring:2019oci}
A.~Behring and W.~Bizoń, \emph{{Higgs decay into massive b-quarks at NNLO QCD
  in the nested soft-collinear subtraction scheme}},
  \href{https://doi.org/10.1007/JHEP01(2020)189}{\emph{JHEP} {\bfseries 01}
  (2020) 189} [\href{https://arxiv.org/abs/1911.11524}{{\ttfamily
  1911.11524}}].

\bibitem{Aaboud:2018xdt}
{\scshape ATLAS} collaboration, \emph{{Measurements of Higgs boson properties
  in the diphoton decay channel with 36 fb$^{-1}$ of $pp$ collision data at
  $\sqrt{s} = 13$ TeV with the ATLAS detector}},
  \href{https://doi.org/10.1103/PhysRevD.98.052005}{\emph{Phys. Rev. D}
  {\bfseries 98} (2018) 052005}
  [\href{https://arxiv.org/abs/1802.04146}{{\ttfamily 1802.04146}}].

\bibitem{Sirunyan:2018sgc}
{\scshape CMS} collaboration, \emph{{Measurement and interpretation of
  differential cross sections for Higgs boson production at $\sqrt{s} =$ 13
  TeV}}, \href{https://doi.org/10.1016/j.physletb.2019.03.059}{\emph{Phys.
  Lett. B} {\bfseries 792} (2019) 369}
  [\href{https://arxiv.org/abs/1812.06504}{{\ttfamily 1812.06504}}].

\bibitem{Chen:2014gva}
X.~Chen, T.~Gehrmann, E.~W.~N. Glover and M.~Jaquier, \emph{{Precise QCD
  predictions for the production of Higgs + jet final states}},
  \href{https://doi.org/10.1016/j.physletb.2014.11.021}{\emph{Phys. Lett.}
  {\bfseries B740} (2015) 147}
  [\href{https://arxiv.org/abs/1408.5325}{{\ttfamily 1408.5325}}].

\bibitem{Boughezal:2015dra}
R.~Boughezal, F.~Caola, K.~Melnikov, F.~Petriello and M.~Schulze, \emph{{Higgs
  boson production in association with a jet at next-to-next-to-leading
  order}}, \href{https://doi.org/10.1103/PhysRevLett.115.082003}{\emph{Phys.
  Rev. Lett.} {\bfseries 115} (2015) 082003}
  [\href{https://arxiv.org/abs/1504.07922}{{\ttfamily 1504.07922}}].

\bibitem{Boughezal:2015aha}
R.~Boughezal, C.~Focke, W.~Giele, X.~Liu and F.~Petriello, \emph{{Higgs boson
  production in association with a jet at NNLO using jettiness subtraction}},
  \href{https://doi.org/10.1016/j.physletb.2015.06.055}{\emph{Phys. Lett.}
  {\bfseries B748} (2015) 5}
  [\href{https://arxiv.org/abs/1505.03893}{{\ttfamily 1505.03893}}].

\bibitem{Chen:2016zka}
X.~Chen, J.~Cruz-Martinez, T.~Gehrmann, E.~W.~N. Glover and M.~Jaquier,
  \emph{{NNLO QCD corrections to Higgs boson production at large transverse
  momentum}}, \href{https://doi.org/10.1007/JHEP10(2016)066}{\emph{JHEP}
  {\bfseries 10} (2016) 066}
  [\href{https://arxiv.org/abs/1607.08817}{{\ttfamily 1607.08817}}].

\bibitem{Chen:2018pzu}
X.~Chen, T.~Gehrmann, E.~W.~N. Glover, A.~Huss, Y.~Li, D.~Neill et~al.,
  \emph{{Precise QCD Description of the Higgs Boson Transverse Momentum
  Spectrum}}, \href{https://doi.org/10.1016/j.physletb.2018.11.037}{\emph{Phys.
  Lett. B} {\bfseries 788} (2019) 425}
  [\href{https://arxiv.org/abs/1805.00736}{{\ttfamily 1805.00736}}].

\bibitem{Chen:2019wxf}
X.~Chen, T.~Gehrmann, E.~W.~N. Glover and A.~Huss, \emph{{Fiducial cross
  sections for the four-lepton decay mode in Higgs-plus-jet production up to
  NNLO QCD}}, \href{https://doi.org/10.1007/JHEP07(2019)052}{\emph{JHEP}
  {\bfseries 07} (2019) 052}
  [\href{https://arxiv.org/abs/1905.13738}{{\ttfamily 1905.13738}}].

\bibitem{Campbell:2019gmd}
J.~M. Campbell, R.~K. Ellis and S.~Seth, \emph{{H + 1 jet production
  revisited}}, \href{https://doi.org/10.1007/JHEP10(2019)136}{\emph{JHEP}
  {\bfseries 10} (2019) 136}
  [\href{https://arxiv.org/abs/1906.01020}{{\ttfamily 1906.01020}}].

\bibitem{Cieri:2018oms}
L.~Cieri, X.~Chen, T.~Gehrmann, E.~W.~N. Glover and A.~Huss, \emph{{Higgs boson
  production at the LHC using the $q_T$ subtraction formalism at N$^3$LO QCD}},
  \href{https://doi.org/10.1007/JHEP02(2019)096}{\emph{JHEP} {\bfseries 02}
  (2019) 096} [\href{https://arxiv.org/abs/1807.11501}{{\ttfamily
  1807.11501}}].

\bibitem{Chetyrkin:1999qi}
K.~Chetyrkin and M.~Steinhauser, \emph{{The Relation between the MS-bar and the
  on-shell quark mass at order alpha(s)**3}},
  \href{https://doi.org/10.1016/S0550-3213(99)00784-1}{\emph{Nucl. Phys. B}
  {\bfseries 573} (2000) 617}
  [\href{https://arxiv.org/abs/hep-ph/9911434}{{\ttfamily hep-ph/9911434}}].

\bibitem{Melnikov:2000qh}
K.~Melnikov and T.~v. Ritbergen, \emph{{The Three loop relation between the
  MS-bar and the pole quark masses}},
  \href{https://doi.org/10.1016/S0370-2693(00)00507-4}{\emph{Phys. Lett. B}
  {\bfseries 482} (2000) 99}
  [\href{https://arxiv.org/abs/hep-ph/9912391}{{\ttfamily hep-ph/9912391}}].

\bibitem{Mondini:2019vub}
R.~Mondini and C.~Williams, \emph{{$ H\to b\overline{b}j $ at
  next-to-next-to-leading order accuracy}},
  \href{https://doi.org/10.1007/JHEP06(2019)120}{\emph{JHEP} {\bfseries 06}
  (2019) 120} [\href{https://arxiv.org/abs/1904.08961}{{\ttfamily
  1904.08961}}].

\bibitem{Gaunt:2015pea}
J.~Gaunt, M.~Stahlhofen, F.~J. Tackmann and J.~R. Walsh, \emph{{N-jettiness
  Subtractions for NNLO QCD Calculations}},
  \href{https://doi.org/10.1007/JHEP09(2015)058}{\emph{JHEP} {\bfseries 09}
  (2015) 058} [\href{https://arxiv.org/abs/1505.04794}{{\ttfamily
  1505.04794}}].

\bibitem{Boughezal:2015dva}
R.~Boughezal, C.~Focke, X.~Liu and F.~Petriello, \emph{{$W$-boson production in
  association with a jet at next-to-next-to-leading order in perturbative
  QCD}}, \href{https://doi.org/10.1103/PhysRevLett.115.062002}{\emph{Phys. Rev.
  Lett.} {\bfseries 115} (2015) 062002}
  [\href{https://arxiv.org/abs/1504.02131}{{\ttfamily 1504.02131}}].

\bibitem{Boughezal:2015ded}
R.~Boughezal, J.~M. Campbell, R.~K. Ellis, C.~Focke, W.~T. Giele, X.~Liu
  et~al., \emph{{Z-boson production in association with a jet at
  next-to-next-to-leading order in perturbative QCD}},
  \href{https://arxiv.org/abs/1512.01291}{{\ttfamily 1512.01291}}.

\bibitem{Campbell:2016lzl}
J.~M. Campbell, R.~K. Ellis and C.~Williams, \emph{{Direct Photon Production at
  Next-to\textendash{}Next-to-Leading Order}},
  \href{https://doi.org/10.1103/PhysRevLett.118.222001}{\emph{Phys. Rev. Lett.}
  {\bfseries 118} (2017) 222001}
  [\href{https://arxiv.org/abs/1612.04333}{{\ttfamily 1612.04333}}].

\bibitem{Stewart:2010tn}
I.~W. Stewart, F.~J. Tackmann and W.~J. Waalewijn, \emph{{N-Jettiness: An
  Inclusive Event Shape to Veto Jets}},
  \href{https://doi.org/10.1103/PhysRevLett.105.092002}{\emph{Phys. Rev. Lett.}
  {\bfseries 105} (2010) 092002}
  [\href{https://arxiv.org/abs/1004.2489}{{\ttfamily 1004.2489}}].

\bibitem{Jouttenus:2011wh}
T.~T. Jouttenus, I.~W. Stewart, F.~J. Tackmann and W.~J. Waalewijn, \emph{{The
  Soft Function for Exclusive N-Jet Production at Hadron Colliders}},
  \href{https://doi.org/10.1103/PhysRevD.83.114030}{\emph{Phys. Rev. D}
  {\bfseries 83} (2011) 114030}
  [\href{https://arxiv.org/abs/1102.4344}{{\ttfamily 1102.4344}}].

\bibitem{Jouttenus:2013hs}
T.~T. Jouttenus, I.~W. Stewart, F.~J. Tackmann and W.~J. Waalewijn, \emph{{Jet
  mass spectra in Higgs boson plus one jet at next-to-next-to-leading
  logarithmic order}},
  \href{https://doi.org/10.1103/PhysRevD.88.054031}{\emph{Phys. Rev. D}
  {\bfseries 88} (2013) 054031}
  [\href{https://arxiv.org/abs/1302.0846}{{\ttfamily 1302.0846}}].

\bibitem{Campbell:2017hsw}
J.~M. Campbell, R.~K. Ellis, R.~Mondini and C.~Williams, \emph{{The NNLO QCD
  soft function for 1-jettiness}},
  \href{https://doi.org/10.1140/epjc/s10052-018-5732-1}{\emph{Eur. Phys. J.}
  {\bfseries C78} (2018) 234}
  [\href{https://arxiv.org/abs/1711.09984}{{\ttfamily 1711.09984}}].

\bibitem{Boughezal:2015eha}
R.~Boughezal, X.~Liu and F.~Petriello, \emph{{$N$-jettiness soft function at
  next-to-next-to-leading order}},
  \href{https://doi.org/10.1103/PhysRevD.91.094035}{\emph{Phys. Rev.}
  {\bfseries D91} (2015) 094035}
  [\href{https://arxiv.org/abs/1504.02540}{{\ttfamily 1504.02540}}].

\bibitem{Becher:2010pd}
T.~Becher and G.~Bell, \emph{{The gluon jet function at two-loop order}},
  \href{https://doi.org/10.1016/j.physletb.2010.11.036}{\emph{Phys. Lett.}
  {\bfseries B695} (2011) 252}
  [\href{https://arxiv.org/abs/1008.1936}{{\ttfamily 1008.1936}}].

\bibitem{Becher:2006qw}
T.~Becher and M.~Neubert, \emph{{Toward a NNLO calculation of the {$\bar{B}$}
  ---> {$X_{s}$} + gamma decay rate with a cut on photon energy. II. Two-loop
  result for the jet function}},
  \href{https://doi.org/10.1016/j.physletb.2006.04.046}{\emph{Phys. Lett.}
  {\bfseries B637} (2006) 251}
  [\href{https://arxiv.org/abs/hep-ph/0603140}{{\ttfamily hep-ph/0603140}}].

\bibitem{Gaunt:2014xga}
J.~R. Gaunt, M.~Stahlhofen and F.~J. Tackmann, \emph{{The Quark Beam Function
  at Two Loops}}, \href{https://doi.org/10.1007/JHEP04(2014)113}{\emph{JHEP}
  {\bfseries 04} (2014) 113} [\href{https://arxiv.org/abs/1401.5478}{{\ttfamily
  1401.5478}}].

\bibitem{Gaunt:2014cfa}
J.~Gaunt, M.~Stahlhofen and F.~J. Tackmann, \emph{{The Gluon Beam Function at
  Two Loops}}, \href{https://doi.org/10.1007/JHEP08(2014)020}{\emph{JHEP}
  {\bfseries 08} (2014) 020} [\href{https://arxiv.org/abs/1405.1044}{{\ttfamily
  1405.1044}}].

\bibitem{Ahmed:2014pka}
T.~Ahmed, M.~Mahakhud, P.~Mathews, N.~Rana and V.~Ravindran, \emph{{Two-loop
  QCD corrections to Higgs $\to b+\overline{b}+g$ amplitude}},
  \href{https://doi.org/10.1007/JHEP08(2014)075}{\emph{JHEP} {\bfseries 08}
  (2014) 075} [\href{https://arxiv.org/abs/1405.2324}{{\ttfamily 1405.2324}}].

\bibitem{Gehrmann:2002zr}
T.~Gehrmann and E.~Remiddi, \emph{{Analytic continuation of massless two loop
  four point functions}},
  \href{https://doi.org/10.1016/S0550-3213(02)00569-2}{\emph{Nucl. Phys. B}
  {\bfseries 640} (2002) 379}
  [\href{https://arxiv.org/abs/hep-ph/0207020}{{\ttfamily hep-ph/0207020}}].

\bibitem{Li:2013lsa}
Y.~Li and H.~X. Zhu, \emph{{Single soft gluon emission at two loops}},
  \href{https://doi.org/10.1007/JHEP11(2013)080}{\emph{JHEP} {\bfseries 11}
  (2013) 080} [\href{https://arxiv.org/abs/1309.4391}{{\ttfamily 1309.4391}}].

\bibitem{Badger:2004uk}
S.~Badger and E.~Glover, \emph{{Two loop splitting functions in QCD}},
  \href{https://doi.org/10.1088/1126-6708/2004/07/040}{\emph{JHEP} {\bfseries
  07} (2004) 040} [\href{https://arxiv.org/abs/hep-ph/0405236}{{\ttfamily
  hep-ph/0405236}}].

\bibitem{Jones:2018hbb}
S.~P. Jones, M.~Kerner and G.~Luisoni, \emph{{Next-to-Leading-Order QCD
  Corrections to Higgs Boson Plus Jet Production with Full Top-Quark Mass
  Dependence}},
  \href{https://doi.org/10.1103/PhysRevLett.120.162001}{\emph{Phys. Rev. Lett.}
  {\bfseries 120} (2018) 162001}
  [\href{https://arxiv.org/abs/1802.00349}{{\ttfamily 1802.00349}}].

\bibitem{Bizon:2021nvf}
W.~Bizon, K.~Melnikov and J.~Quarroz, \emph{{On the interference of $ggH$ and
  $c\bar{c}H$ Higgs production mechanisms and the determination of charm Yukawa
  coupling at the LHC}},  \href{https://arxiv.org/abs/2102.04242}{{\ttfamily
  2102.04242}}.

\bibitem{Mondini:2020uyy}
R.~Mondini, U.~Schubert and C.~Williams, \emph{{Top-induced contributions to
  $H\rightarrow b\bar{b}$ and $H\rightarrow c\bar{c}$ at
  $\mathcal{O}(\alpha_s^3)$}},
  \href{https://doi.org/10.1007/JHEP12(2020)058}{\emph{JHEP} {\bfseries 12}
  (2020) 058} [\href{https://arxiv.org/abs/2006.03563}{{\ttfamily
  2006.03563}}].

\bibitem{Campbell:1999ah}
J.~M. Campbell and R.~K. Ellis, \emph{{An Update on vector boson pair
  production at hadron colliders}},
  \href{https://doi.org/10.1103/PhysRevD.60.113006}{\emph{Phys. Rev.}
  {\bfseries D60} (1999) 113006}
  [\href{https://arxiv.org/abs/hep-ph/9905386}{{\ttfamily hep-ph/9905386}}].

\bibitem{Campbell:2011bn}
J.~M. Campbell, R.~K. Ellis and C.~Williams, \emph{{Vector boson pair
  production at the LHC}},
  \href{https://doi.org/10.1007/JHEP07(2011)018}{\emph{JHEP} {\bfseries 07}
  (2011) 018} [\href{https://arxiv.org/abs/1105.0020}{{\ttfamily 1105.0020}}].

\bibitem{Campbell:2015qma}
J.~M. Campbell, R.~K. Ellis and W.~T. Giele, \emph{{A Multi-Threaded Version of
  MCFM}}, \href{https://doi.org/10.1140/epjc/s10052-015-3461-2}{\emph{Eur.
  Phys. J.} {\bfseries C75} (2015) 246}
  [\href{https://arxiv.org/abs/1503.06182}{{\ttfamily 1503.06182}}].

\bibitem{Boughezal:2016wmq}
R.~Boughezal, J.~M. Campbell, R.~K. Ellis, C.~Focke, W.~Giele, X.~Liu et~al.,
  \emph{{Color singlet production at NNLO in MCFM}},
  \href{https://doi.org/10.1140/epjc/s10052-016-4558-y}{\emph{Eur. Phys. J.}
  {\bfseries C77} (2017) 7} [\href{https://arxiv.org/abs/1605.08011}{{\ttfamily
  1605.08011}}].

\bibitem{Harlander:2003ai}
R.~V. Harlander and W.~B. Kilgore, \emph{{Higgs boson production in bottom
  quark fusion at next-to-next-to leading order}},
  \href{https://doi.org/10.1103/PhysRevD.68.013001}{\emph{Phys. Rev. D}
  {\bfseries 68} (2003) 013001}
  [\href{https://arxiv.org/abs/hep-ph/0304035}{{\ttfamily hep-ph/0304035}}].

\bibitem{Harland-Lang:2014zoa}
L.~Harland-Lang, A.~Martin, P.~Motylinski and R.~Thorne, \emph{{Parton
  distributions in the LHC era: MMHT 2014 PDFs}},
  \href{https://doi.org/10.1140/epjc/s10052-015-3397-6}{\emph{Eur. Phys. J. C}
  {\bfseries 75} (2015) 204} [\href{https://arxiv.org/abs/1412.3989}{{\ttfamily
  1412.3989}}].

\bibitem{Chetyrkin:2000yt}
K.~G. Chetyrkin, J.~H. Kuhn and M.~Steinhauser, \emph{{RunDec: A Mathematica
  package for running and decoupling of the strong coupling and quark masses}},
  \href{https://doi.org/10.1016/S0010-4655(00)00155-7}{\emph{Comput. Phys.
  Commun.} {\bfseries 133} (2000) 43}
  [\href{https://arxiv.org/abs/hep-ph/0004189}{{\ttfamily hep-ph/0004189}}].

\bibitem{Catani:1996vz}
S.~Catani and M.~H. Seymour, \emph{{A General algorithm for calculating jet
  cross-sections in NLO QCD}},
  \href{https://doi.org/10.1016/S0550-3213(96)00589-5}{\emph{Nucl. Phys.}
  {\bfseries B485} (1997) 291}
  [\href{https://arxiv.org/abs/hep-ph/9605323}{{\ttfamily hep-ph/9605323}}].

\bibitem{Nagy:1998bb}
Z.~Nagy and Z.~Trocsanyi, \emph{{Next-to-leading order calculation of four jet
  observables in electron positron annihilation}},
  \href{https://doi.org/10.1103/PhysRevD.62.099902}{\emph{Phys. Rev. D}
  {\bfseries 59} (1999) 014020}
  [\href{https://arxiv.org/abs/hep-ph/9806317}{{\ttfamily hep-ph/9806317}}].

\bibitem{Moult:2016fqy}
I.~Moult, L.~Rothen, I.~W. Stewart, F.~J. Tackmann and H.~X. Zhu,
  \emph{{Subleading Power Corrections for N-Jettiness Subtractions}},
  \href{https://doi.org/10.1103/PhysRevD.95.074023}{\emph{Phys. Rev. D}
  {\bfseries 95} (2017) 074023}
  [\href{https://arxiv.org/abs/1612.00450}{{\ttfamily 1612.00450}}].

\bibitem{Dulat:2015mca}
S.~Dulat, T.~J. Hou, J.~Gao, M.~Guzzi, J.~Huston, P.~Nadolsky et~al.,
  \emph{{The CT14 Global Analysis of Quantum Chromodynamics}},
  \href{https://arxiv.org/abs/1506.07443}{{\ttfamily 1506.07443}}.

\bibitem{Buckley:2014ana}
A.~Buckley, J.~Ferrando, S.~Lloyd, K.~Nordstr\"om, B.~Page, M.~R\"ufenacht
  et~al., \emph{{LHAPDF6: parton density access in the LHC precision era}},
  \href{https://doi.org/10.1140/epjc/s10052-015-3318-8}{\emph{Eur. Phys. J. C}
  {\bfseries 75} (2015) 132} [\href{https://arxiv.org/abs/1412.7420}{{\ttfamily
  1412.7420}}].

\end{thebibliography}\endgroup

\end{document}